\documentclass[reprint,amsmath,amssymb,superscriptaddress,prc,floatfix]{revtex4-1}
\usepackage{amsmath}
\usepackage{graphicx}
\usepackage{longtable}
\usepackage{bm}
\usepackage[caption=false,subrefformat=parens]{subfig}
\usepackage{hyperref}
\hypersetup{colorlinks=true,breaklinks,linkcolor=red,citecolor=blue}

\begin{document}

\title{Fission dynamics within time-dependent Hartree-Fock: deformation-induced fission}

\author{Philip Goddard}
\affiliation{Department of Physics, Faculty of Engineering and Physical Sciences, University of Surrey, Guildford, Surrey GU2 7XH, United Kingdom}

\author{Paul Stevenson}
\email{p.stevenson@surrey.ac.uk}
\affiliation{Department of Physics, Faculty of Engineering and Physical Sciences, University of Surrey, Guildford, Surrey GU2 7XH, United Kingdom}

\author{Arnau Rios}
\email{a.rios@surrey.ac.uk}
\affiliation{Department of Physics, Faculty of Engineering and Physical Sciences, University of Surrey, Guildford, Surrey GU2 7XH, United Kingdom}

\date{\today}

\begin{abstract}
\begin{description} 
\item[Background] Nuclear fission is a complex large-amplitude collective decay mode in heavy nuclei.  Microscopic density functional studies of fission have previously concentrated on adiabatic approaches based on constrained static calculations ignoring dynamical excitations of the fissioning nucleus and the daughter products.
\item[Purpose] We explore the ability of dynamic mean-field methods to describe fast fission processes beyond the fission barrier, using the nuclide $^{240}$Pu as an example.
\item[Methods] Time-dependent Hartree-Fock calculations based on the Skyrme interaction are used to calculate nonadiabatic fission paths, beginning from static constrained Hartree-Fock calculations.  The properties of the dynamic states are interpreted in terms of the nature of their collective motion.  Fission product properties are compared to data.
\item[Results] Parent nuclei constrained to begin dynamic evolution with a deformation less than the fission barrier exhibit giant-resonance-type behaviour.  Those beginning just beyond the barrier explore large amplitude motion but do not fission, whereas those beginning beyond the two-fragment pathway crossing fission to final states which differ according to the exact initial deformation.
\item[Conclusions] Time-dependent Hartree-Fock is able to give a good qualitative and quantitative description of fast fission, provided one begins from a sufficiently deformed state.
\end{description}
\end{abstract}


\maketitle

\section{Introduction}

Studies of nuclear fission have been ongoing since the discovery of the process in 1938 by Hahn and Strassmann \cite{Hah39}. The actinide nuclide $^{240}$Pu has long been a case of interest, as spontaneous fission presents itself as a decay mechanism with significant probability, relative to other isotopes in the actinide region. This allows for quantitative comparisons between spontaneous and induced fission \cite{Thi81,Wag84,Tor71,Wat62}. Experimentally, fission can be induced by a variety of techniques, including neutron-induced fission, fission induced by more complex projectiles, and photo-fission \cite{Thi81,Hof74}. Recent experimental campaigns have investigated $\beta$-delayed fission \cite{And13}.

\begin{figure}[tb]
\begin{center} 
\includegraphics[width=\linewidth]{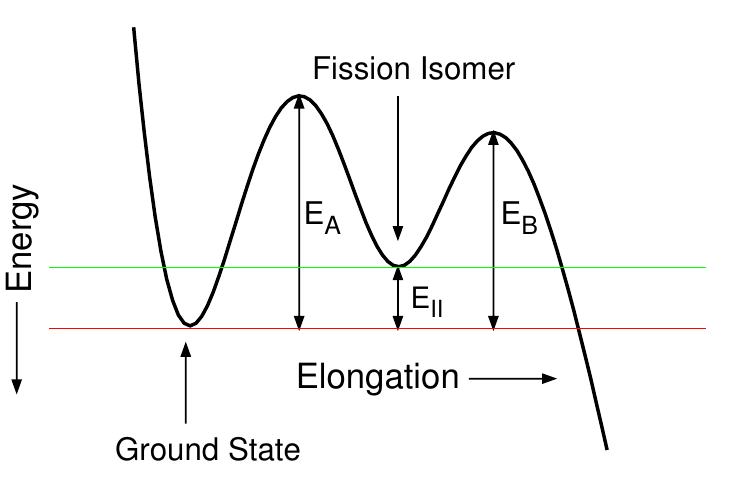} 
\caption{(Color online) Schematic of typical potential energy surface obtained when increasing the elongation (which corresponds to quadrupole deformation) of an actinide nucleus. The energies $E_A$, $E_B$ and $E_{II}$ relate to properties of the fission barriers, and correspond to the definitions in Ref.\@ \cite{Mol09}.}
\label{humps}
\end{center}
\end{figure}

Theoretically, microscopic studies have focused upon the role of the quadrupole degree of freedom in forming the fission pathway, as exemplified by constrained mean-field calculations \cite{Flo73,Flo74}. The typical observed behaviour in actinide nuclei for the binding energy as a function of increasing quadrupole deformation is to follow a multi-humped pathway (see Fig.\@ \ref{humps}). When considering the potential energy surface (PES), starting from the ground state, then increasing the quadrupole deformation will result in a first fission barrier. By increasing the deformation further, a secondary minimum, corresponding to an isomer, is found. Beyond this minimum, a second fission barrier is encountered, and past this barrier, the general consensus is that it becomes more energetically favourable for the nucleus to fission. The energies $E_A$, $E_B$ and $E_{II}$ presented in Fig.\@ \ref{humps} correspond to those defined in Ref.\@ \cite{Mol09}: the energy difference between the ground state and the peak of the first fission barrier, the difference between the ground state and the peak of the second fission barrier, and the difference between the ground state and fission isomer, respectively. In some exotic cases, triple-humped potential surfaces are expected \cite{Mol72,Gav76,Bjo80,McD13}. Although the multi-humped behaviour of the energy surface cannot be measured directly, experimental evidence points towards this characteristic structure \cite{Spe74,Bjo80}. 

For fission studies, the quadrupole degree of freedom is of capital importance, as it describes the elongation of the nucleus \cite{gre96}. Additionally, as many nuclei are observed to fission with asymmetric mass distributions, the octupole degree of freedom is vital to describe any mass reflection asymmetry. Modern density functional theory (DFT) solvers are able to perform symmetry-unrestricted calculations which allow, in principle, any and multiple degrees of freedom to be explored \cite{Sch12}. As well as considering deformation degrees of freedom, constraints can be imposed from alternate perspectives to study fission in static calculations. Some studies, for instance, assume a symmetric fission fragment path \cite{Sim13} or a constrained multi-configurational static solution before investigating time evolution \cite{Gou05}. In fact, time-dependent generator coordinate method calculations provide a generalization of the approach presented here to take into account the evolution of collective coordinates \cite{Bernard2011}.

The approach of calculating the PES to describe fission, regardless of the number of particular degrees of freedom constrained, is limited to producing a series of static solutions which attempt to describe a dynamic process, resulting in an effectively adiabatic approximation. Shape-constrained DFT calculations produce Slater determinants which contain no internal excitations. Some attempts have been made to account for finite-temperature effects as sources of dissipation \cite{McD14a}. How such effects would couple with the nonadiabatic time-dependent Hartree-Fock (TDHF) approach is an open question that deserves further study \cite{Luo1999}. Time-dependent techniques coupled with constrained calculations may therefore yield new, insightful results, because they describe the dynamics of a fissioning system. Time-dependent Hartree-Fock \cite{Dir30} presents itself as a candidate method, as it is able to temporally evolve Slater determinants which begin as a solution to the constrained static Hartree-Fock (HF) equations.

The calculational basis for this study is the so-called TDHF technique. We note, however, that nowadays this should probably be called ``time-dependent density functional", because it is based upon a nuclear energy density functional.  The slightly less correct TDHF moniker has stuck within the nuclear physics literature, though, and we keep this convention in the present work.

TDHF is the basic lowest-order microscopic dynamical mean-field theory, first proposed by Dirac \cite{Dir30} and later on applied to more or less realistic nuclear systems in the 1970s$-$1980s \cite{Eng75,Bonche1976,Neg78,Str79,Uma85}.  A practical implementation involves beginning from an energy density functional and using the variational principle to obtain Hartree-Fock-like equations for the static initial state. The time evolution equations, based on the same energy density functional, are then run on this initial state. We use the Skyrme energy density functional, depending on the local densities and currents,
\begin{equation}
  E=E_{\mathrm sky}(\rho,\bm{J},\tau,\bm{s},\bm{j},\xi)
\end{equation}
where $\rho$ is the particle density, $\bm{J}$ is the vector part of the spin-current tensor, $\tau$ is the kinetic density, $\bm{s}$ is the spin density, $\bm{j}$ is the particle current and $\xi$ is the pairing density \cite{Ben03}.  These densities and currents include time-odd fields ($\bm{s}$ and $\bm{j}$), which are only active in the dynamic part of the calculation.  Only those time-odd fields that couple to the necessary time-even fields through Galilean invariance are included in the present calculation.  Full details of the practical aspects of solving the static and dynamic HF equations, along with the explicit details of the density functional are contained in the documentation of the {\sc sky3d} code \cite{Mar13}.

There have been several historic attempts to describe fission dynamics using TDHF. The pioneering attempt by Negele \emph{et al.}~\cite{Neg78} was followed by a series of studies that suffered from computational power limitations and were hindered by axial symmetry and restricted forms of the nucleon-nucleon interaction~\cite{Oko83,Jun88}. With a new generation of TDHF solvers able to perform symmetry-unrestricted, three-dimensional calculations \cite{Mar13,Kim97,Uma91}, modern TDHF studies have begun to take a renewed interest in fission \cite{Uma10,Sim13,Scamps2015}. Here we use a modified version of the recently published code {\sc sky3d} \cite{Mar13}, with shape constraints explicitly included. We then solve the HF + BCS equations in a three-dimensional Cartesian basis and evolve the calculated states using TDHF.
 
Spontaneous fission cannot be accessed directly within TDHF. To reach a fissioned configuration from the ground or isomeric state, the nucleus must tunnel through the barriers in the PES. While TDHF allows a quantum mechanical description of single-particle wave functions, the collective motion is semiclassical, hence forbidding tunnelling in collective coordinates \cite{rin80}. In contrast, TDHF is  suitable for exploring the dynamics of induced fission. The potential challenging issue is the incorporation of the fissioning mechanism within the TDHF framework. Here we follow the strategy of finding constrained Hartree-Fock (CHF) states and use them as initial conditions in a TDHF calculation. Our aim is to investigate how the underlying deformed structure pushes the parent nucleus towards fissioning paths, if it does so at all. 

We also run TDHF simulations from initial states that lie past the fission barrier. At that stage, one is somehow mimicking spontaneous decay in the sense that these represent the states right after tunnelling. The complex quantum dynamics that occur in the lead-up to the point at which we let the TDHF calculations take over will clearly populate a range of different configurations, leading to a spread of fission products.  Our dynamical calculations only take a handful of initial states on a single constrained quadrupole PES, as a proxy for the quantum nature of the ``tunnelling'' process.  A more realistic way of sampling may be to find multiple PES solutions at the same energy and start from those.  We note that our different initial configurations correspond to slightly different total energy content in the system.

The paper is organised as follows. In Sec.~\ref{sec:static}, the static constrained starting points are discussed. Section \ref{sec:dynamic} explores the dynamics of the fissioning nucleus, while Sec.~\ref{sec:fragments} looks in more detail at the dynamics of the fission fragments.  Some concluding remarks are given in Sec.~\ref{sec:conclusion}.

\section{Static configurations in $^{240}$Pu}
\label{sec:static}

We begin the investigation of fission by examining the PES for a nucleus of interest. The reproduction of the double-humped fission barriers of actinide nuclei are often used as a benchmark test for nuclear models \cite{Sto07,Bertsch2015}. Owing to the wealth of data available from experimental \cite{Thi81,Wag84,Tor71,Wat62} and theoretical \cite{Str67,Flo74,Rod14,Mol09} studies, $^{240}$Pu presents itself as a strong candidate for a benchmark test of TDHF to investigate induced nuclear fission \cite{Bertsch2015}. We note, in particular, that configuration mixing plays a relatively small effect in this isotope \cite{Bender2004}. The static quadrupole-constrained PES is firstly calculated to provide a selection of initial states for time evolution. These are obtained by performing an energy minimisation with respect to constraints imposed upon the quadrupole shape degree of freedom. 

Modern symmetry-unrestricted DFT solvers have extended the PES for multiple constraints, for example simultaneously constraining the quadrupole and octupole degrees of freedom to explore two-dimensional deformation surfaces \cite{Sta10}. The approach of calculating a multi-dimensional PES to describe fission within a microscopic framework has enjoyed much recent attention \cite{Sta10,War12,McD13,McD14a,Sch14a}. 
Alternatively, one can explore fission pathways via shell-corrected macroscopic liquid drop models. This technique has been applied to perform exhaustive topographical surveys of deformation space to deduce fission properties of static configurations \cite{Mol01,Mol00,Mol04,Mol09}. However, this approach, regardless of the number of dimensions, is limited to producing a series of static solutions to describe a dynamic process, thus being equivalent to adiabatic motion.  Time-dependent simulations describe the fissioning system, while allowing for internal excitations. This is important for the final, fast stages of fission, which we explore in the present work.

To build shape-constrained ground states, we begin from an arbitrarily deformed state and use the augmented Lagrangian method \cite{Sta10,Noc06} to constrain the quadrupole deformation.  Our purpose is not to pursue an in-depth investigation of multiple shape-constraints in HF calculations, but rather to use the technique to produce initial states to then investigate their time evolution. All other degrees of freedom are assumed to settle into the configuration of minimum energy \cite{rin80}. 

For the results presented here, where only one constraint is applied upon the quadrupole degree of freedom, a masking procedure has been adopted. This limits the space which the nuclear wave functions can explore, allowing a single fission pathway to be explored where the nuclear shape gradually evolves, rather than abruptly jumping between competing energy minima. Details of the masking procedure are discussed in Ref.~\cite{Goddard14}.  Our emphasis is on the fast fission dynamics beyond the barrier, but it may yet be fruitful to use other constraints to generate more starting points. 

\begin{table}[tbh]
\caption{Summary of ground-state and isomer properties of $^{240}$Pu, calculated using {\sc sky3d} with the SkM$^{\ast}$ Skyrme effective interaction. Further details of the calculations are included in the text. Both the ground state and isomer are prolate deformed and axially symmetric.} 
\centering
\begin{tabular}{|c|c|c|c|c|c|c|c|} 
\hline 
 Nucleus & Binding Energy& rms Radius & $\beta_{20}$ &$\beta_{30}$& $\beta_{40}$ \\
 &[MeV]&[fm]&&& \\
\hline\hline 
$^{240}$Pu&-1781.95&5.941&0.280&0.000&0.255 \\
$^{240}$Pu$^\ast$&-1778.91&6.418&0.682&0.000&0.547\\
\hline 
\end{tabular}
\label{tabbbpu}
\end{table}

The ground-state and CHF calculations are performed using the SkM$^\ast$ effective interaction. The fission barrier properties of $^{240}$Pu were considered when fitting the SkM$^\ast$ effective interaction \cite{Bar82}. We perform our static calculations in a regularly spaced Cartesian grid of $40\times40\times40$ points, ranging from $-19.5$ to $19.5$ fm in the $x$, $y$ and $z$ directions. This rather coarse grid gives surprisingly good results, with, e.g., binding energies typically differing by parts in around $10^4$ relative to much finer grids \cite{Mar13}. BCS pairing is included within the static calculation, using the Volume-$\Delta$ interaction \cite{Mar13}, with $184$ neutron and $126$ proton single-particle wave functions and pairing strengths $V_{0,n}=258.96201$ MeV and $V_{0,p}=270.08200$ MeV for neutrons and protons, respectively. The PES for $^{240}$Pu presents a prominent local minimum, corresponding to a fission isomer. If the initial test wave functions (harmonic oscillator states in the case of {\sc sky3d}) are chosen to be prolate with similar deformations, shape unconstrained static calculations converge directly into the isomeric state. This provides two initial points for the CHF calculations, starting at either the ground or the isomeric state. Some properties of the ground state and isomer are presented in Table \ref{tabbbpu}. These compare well with previous results in the literature \cite{Flo74}.  We consider thus that our choice of pairing interaction is suitable. We do not take into account pairing correlations in the dynamical evolution, and hence we do not perform a systematic study of the role of the pairing interaction.

\begin{figure*}[tbh]
\begin{center} 
\includegraphics[width=0.8\linewidth]{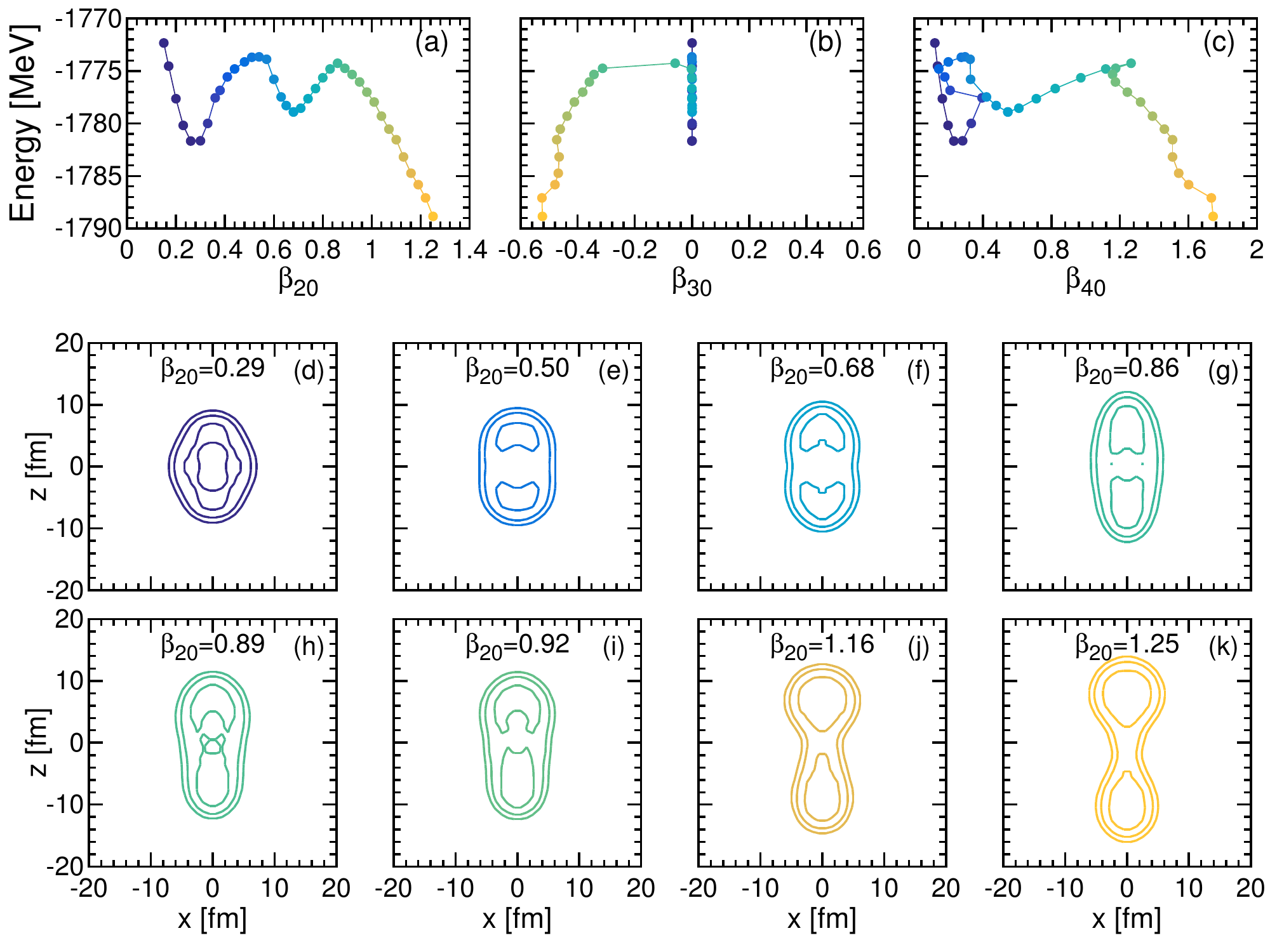} 
\caption{(Color online) Resulting PES for $^{240}$Pu following a constraint of the quadrupole deformation parameter $\beta_{20}$. The top panels display the dependence of the constrained energy on the (a) quadrupole, (b) octupole and (c) hexadecapole moments. Once the second barrier is overcome, a significant octupole deformation (corresponding to mass asymmetry) develops. Bottom panels (d)-(k) display two-dimensional (2D) slices of the 3D density for increasing quadrupole deformation. The isolines correspond to 0.05 particles/fm$^{3}$. It is interesting to note that, for states far beyond the second barrier, scission has not yet occurred.}
\label{pu-esurf-pretty}
\end{center}
\end{figure*} 

One-dimensional projections of the quadrupole-constrained PES for $^{240}$Pu are shown in the top three panels of Fig.\@ \ref{pu-esurf-pretty}. Panel (a) focuses on the dependence of the energy on the constrained quadrupole degree of freedom. Panels (b) and (c) give the corresponding octupole and hexadecapole coordinates associated with each quadrupole configuration. For guidance, the color scheme is chosen to display changing quadrupoles. Two fission barriers are found in the quadrupole degree of freedom (top left panel), peaking at $\beta_{20} \approx 0.50$ and $0.86$ respectively. The ground and isomeric states correspond to the two minima in the PES next to these barriers. 

Table \ref{comparison-pu} presents a comparison between various features of the PES calculated in this work and previous literature. The table also contains measurements of the fission barrier heights and energy differences between the ground and the isomeric states, as defined in Fig.\@ \ref{humps}. We compare our calculations to recent Hartree-Fock-Bogoliubov (HFB) calculations using the Gogny interaction \cite{Rod14}, to older HF calculations employing the Skyrme SIII parametrization \cite{Flo74}, and to macroscopic-microscopic calculations based on the shell-corrected Finite Range Liquid-Drop Model (FRLDM) \cite{Mol09}. There is a general agreement between the barrier geometries in the theoretical methods. The older SIII calculations predict a large second barrier height, most likely because of the assumption of axial symmetry. HF calculations yield barriers higher than either the experimental or the macroscopic-microscopic predictions, perhaps caused by a lack of dynamical effects.  The details of the barrier shapes should be sufficiently well reproduced in the present calculations to deal with the fast fission dynamics beyond the second barrier, as well as to give a qualitative description of the between-barrier dynamics.

\begin{table*}[tbh]
\begin{tabular}{|c|c|c|c|c|} 
\hline
 $E_A$ & $E_B$& $E_{II}$ & Method &Reference  \\
 $[$MeV$]$&$[$MeV$]$&$[$MeV$]$&& \\
\hline\noalign{\smallskip}\hline
8.25&7.68&3.04&HF+BCS, Skyrme SkM$^\ast$ &This work \\
8&13&4&HF+BCS, Skyrme SIII &Table 2 of Ref.\@ \cite{Flo74} \\
9.30&8.40&3.10&HFB, Gogny D1M & Fig.\@ 5 of Ref.\@ \cite{Rod14} \\
5.99&4.91&2.94&Shell-Corrected FRLDM & Table I of Ref.\@ \cite{Mol09}\\ 
6.1$\pm$0.3&6.0$\pm$0.50&2.1$\pm$0.6&Experiment &Fig.\@ 27 of Ref.\@ \cite{Mol09} (Madland)  \\
5.6$\pm$0.2&5.1$\pm$0.20&2.4$\pm$0.3&Experiment &Fig.\@ 27 of Ref.\@ \cite{Mol09} (Madland) \\
\hline 
\end{tabular}
\caption{\label{comparison-pu}
Comparison of properties of the fission barrier for $^{240}$Pu from different calculations (defined in Fig.\@ \ref{humps}). In addition to our work, we present the calculations of Flocard \textit{et al.} \cite{Flo74}, Rodr{\'i}guez-Guzm{\'a}n and Robledo \cite{Rod14}, M{\"o}ller \textit{et al.} \cite{Mol09} and the experimentally inferred data presented in Ref. \cite{Mol09}.
}
\end{table*}

\begin{figure}[tbh]
\begin{center} 
\includegraphics[width=0.7\linewidth]{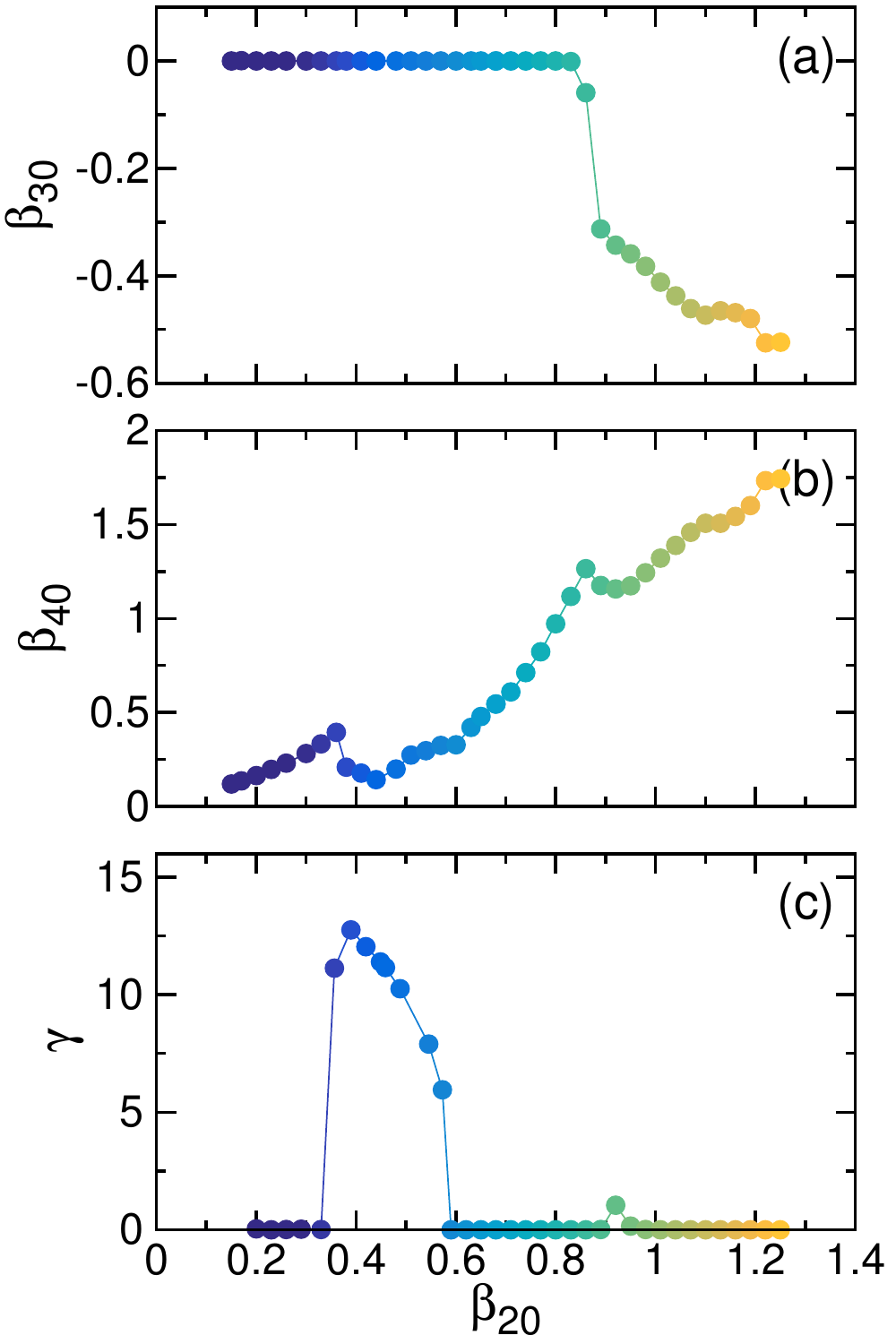} 
\caption{ 
(Color online) (a) Octupole deformation as a function of quadrupole deformation and (b) hexadecapole deformation as a function of quadrupole deformation for the calculated PES of $^{240}$Pu. The octupole deformation, corresponding to a mass asymmetry, rapidly onsets after the second fission barrier is passed. Panel (c) displays the corresponding $\gamma$ deformation parameter. }
\label{qvo}
\end{center}
\end{figure}

Panel (b) of Fig.~\ref{pu-esurf-pretty} shows a prominent octupole deformation setting in at the second fission barrier, as would be expected \cite{Sto07}. The relationship between the quadrupole and octupole deformation parameters for the configurations along the PES is explored in Fig.\@ \ref{qvo} (top panel). Although the calculations have been performed constraining only one deformation degree of freedom, the observed behaviour is typical for the optimum static fission pathway obtained in quadrupole-octupole-constrained deformation surfaces calculated using DFT \cite{Sta10,McD13}. Beyond the second fission barrier, octupole degrees of freedom are explored. Panel (b) of Fig.\@  \ref{qvo} displays the relationship between the quadrupole and hexadecapole deformation parameters. Near the peaks of the first and second barriers, the hexadecapole deformation sharply drops, and recovers subsequently. This corresponds to a transitioning shape as the neck region of the nucleus thins. 

The 3D calculations verify that triaxiality is explored at the first fission barrier. The bottom panel of Fig.\@ \ref{qvo} shows unambiguously a region of nonzero values of $\gamma$. Access to these additional degrees of freedom lowers the calculated barrier height with respect to axially symmetric calculations \cite{Sto07}. We note that triaxiality is explored significantly in the range $0.36\le\beta_{20}\le0.59$, but it is virtually negligible elsewhere.

The slices of the density in the lower panels of Fig.\@ \ref{pu-esurf-pretty} display an increasingly deformed shape as the quadrupole degree of freedom grows. Interestingly, the nucleus has not fissioned in the range of $\beta_{20}$ considered, even for states beyond the second fission barrier. Despite the emergence of a competing fission pathway, a large selection of increasingly deformed states have been obtained. These are useful as starting points for our time-dependent calculations. We note that our configurations explore a range of states, from configurations with a quadrupole deformation less than that of the global HF minimum, to configurations well beyond the second fission barrier.

\begin{figure*}[tbh]
\begin{center} 
\includegraphics[width=0.8\linewidth]{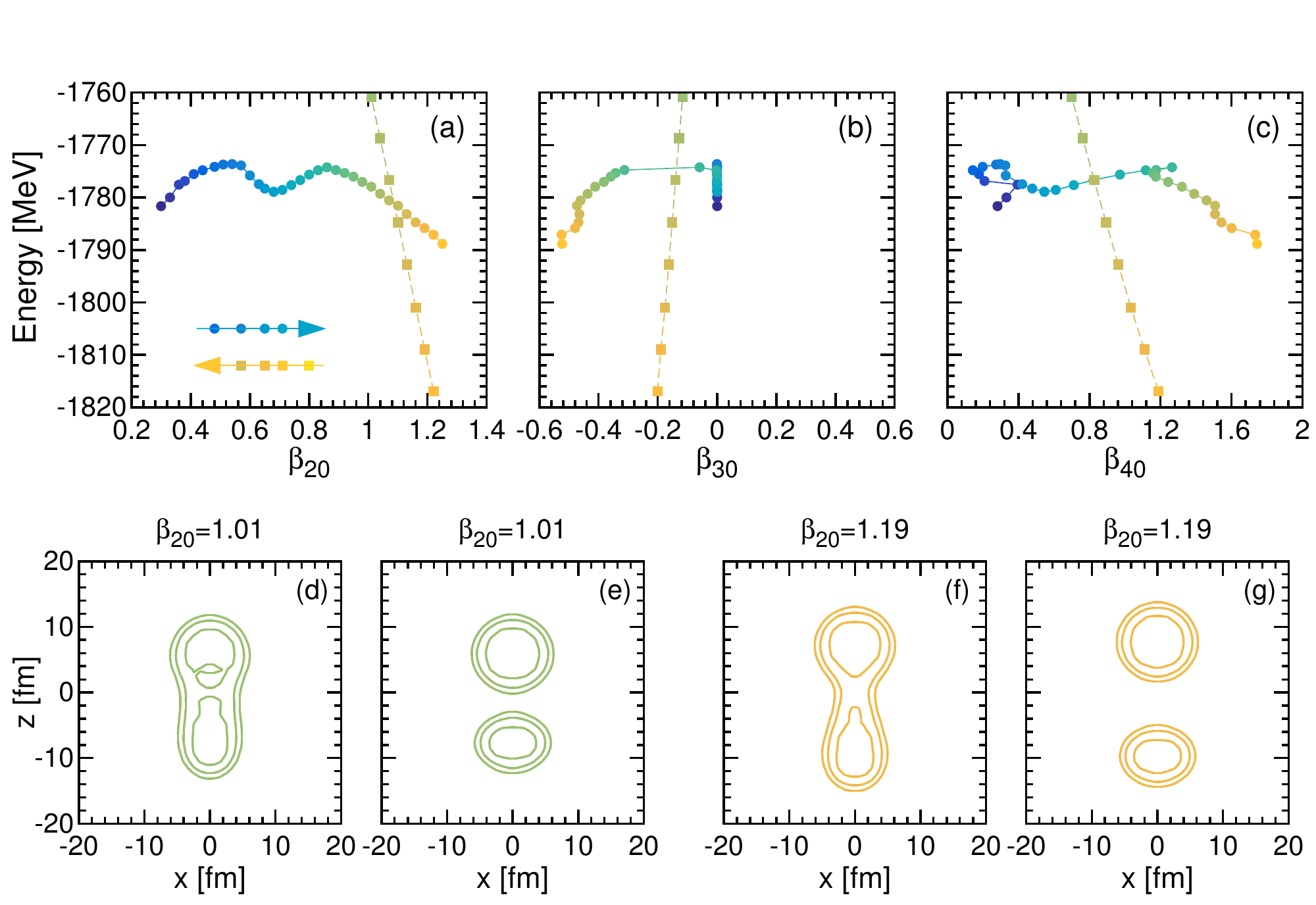} 
\caption{(Color online) Panels (a)$-$(c): One-fragment (solid circles) and two-fragment (solid squares) fission pathways for $^{240}$Pu. The arrows in panel (a) show the direction in which the PES is explored. Beyond $\beta_{20}=1.25$ the one-fragment pathway jumps into the two-fragment pathway, and this state is used as the initial configuration for investigating the latter pathway. Sample density slices on the competing pathways with the same $\beta_{20}$ are shown in the panels (d)-(e) and (f)-(g). The isolines are separated by 0.05 particles/fm$^{3}$. } 
\label{pu-compete}
\end{center}
\end{figure*}

When performing CHF calculations, we noticed that beyond a quadrupole deformation of $\beta_{20}=1.25$, the configuration jumped abruptly to a competing two-fragment fission pathway. This behaviour, owing to the numerics as the calculations converged, proved to be unavoidable. This can be explained by considering the density slices presented in Fig.\@ \ref{pu-compete}. The masking procedure that is implemented in the calculations around the one-fragment configuration does not inhibit a transition to the two-fragment configuration, as the two-fragment configuration fits inside the one-fragment masking region.

The competing two-fragment fission pathway was explored, starting from the state after the calculations jumped pathways. From this configuration, the deformation was incrementally reduced. This competing pathway is shown in Fig.\@ \ref{pu-compete} with solid squares, and may be compared to the original, one-fragment pathway in solid circles. Once the quadrupole deformation parameter is reduced below $\beta_{20}=1.01$, the HF minimum jumps back onto the original fission pathway. The one-fragment pathway is also sometimes denoted as fission valley, whereas the two-fragment pathway \cite{Egido2000,Rod14} has also been referred to as the fusion valley \cite{Berger1990,Bonneau2006}.

The competing pathway, referred to hereafter as the two-fragment pathway \cite{Rod14}, displays remarkably different configurations to that of the one-fragment pathway. Even with identical quadrupole deformations [see panels (d) to (g)], the octupole and hexadecapole deformations and total energy differ significantly. It is exactly this behaviour that the authors of Refs. \cite{Mol09} and \cite{Mol04} identify as a flaw when using CHF to explore the PES. Here, we exploit this feature to gain an insight on the competing fission pathway without having to include a higher number of constraints in the CHF calculations.

The fragments in the two-fragment pathway do not have an integer particle number. For example, for the case of $\beta_{20}=1.19$, the fragments have $A_1=107.14$, $Z_1 =  43.14$, and $A_2=132.85$, $Z_2 = 50.85$. We note, however, that all the fragments in the two-fragment pathway correspond, to the nearest integer particle number, to $^{107}_{43}$Tc and $^{133}_{51}$Sb. It would be instructive to project the individual fragments onto a good particle number \cite{Sim10}, because this would give access to a mass distribution. Our focus here is on the dynamics, though, so we postpone this for future work.

\section{Time evolution of constrained Hartree-Fock states}
\label{sec:dynamic}

The time evolution of the CHF states obtained for $^{240}$Pu may be investigated using the TDHF method. This analysis will focus on the states on the one-fragment fission pathway, starting from configurations beyond the fission isomer (that is, those with $\beta_{20} > 0.68$). We investigate the effect of releasing the imposed shape constraints and time evolving the constrained static states, to gain an insight of the deformation-induced fission (DIF) process in TDHF.

Our TDHF calculations are performed in a larger grid than that used to calculate the initial (static) state.  In most calculations, the Slater determinant solution to the CHF problem is placed in a grid of dimension ($42^3$) points, ranging from $-20.5$ to $20.5$ fm in the $x$, $y$, and $z$ directions. The BCS occupations associated with each wave function remain frozen throughout the dynamical simulation. This approximation is relatively harsh, but provides a substantial computational advantage. Our study represents a preliminary attempt to gauge the potential of TDHF techniques in pre- and post-fission dynamics. As such, while time-dependent pairing effects are likely to be relevant for quantitative predictions, we stay at a relatively qualitative level. Comparisons with observables should therefore be considered with care. We refer the reader to Refs.~\cite{Bl'ocki1976,Ste11,Scamps2015} for details on more thorough treatments of dynamical pairing effects. 

During our calculations, nuclear fragments above particle emission threshold will be created.  Some single particle wave functions will thus be free to explore the entire space of the calculation, up to the box boundaries.  Our periodic boundary conditions thus may cause artificial effects, especially in the analysis of fragment vibrations.  While spherical TDHF calculations can be performed in an analytic continuum \cite{Par13,Par14}, the available numerical methods that can be applied to 3D cases  are computationally expensive and may not be suitable for large amplitude processes \cite{Nak01,Ued03}. When analysing nuclear dynamics, we apply a spatial mask to ensure that observables correspond only to the nuclei and not dripped particles. For separated fragments, individual co-moving masks are used. Further details can be found in Ref.~\cite{Goddard14}.

As one increases the $\beta_{20}$ deformation of the $t=0$ state, different dynamics take place.  Before the second barrier, no fission can occur without collective tunnelling not open to the TDHF method. For these initial configurations, we see large-amplitude collective motion, which has its own interest \cite{Goddard14}. This suggests that the TDHF wave functions are exploring a local minimum in multidimensional deformation space and that there is an inhibition in rearranging substantially the nuclear density while keeping the total energy constant. As a matter of fact, the corresponding power spectra are in qualitative agreement to those obtained in a giant-resonance calculation. We do not explore these issues further here, as our emphasis is on final fissioning states.

\subsection{States Between the One- and the Two-Fragment Pathways}

\begin{figure*}[tbh]
\begin{center} 
\includegraphics[width=0.7\linewidth]{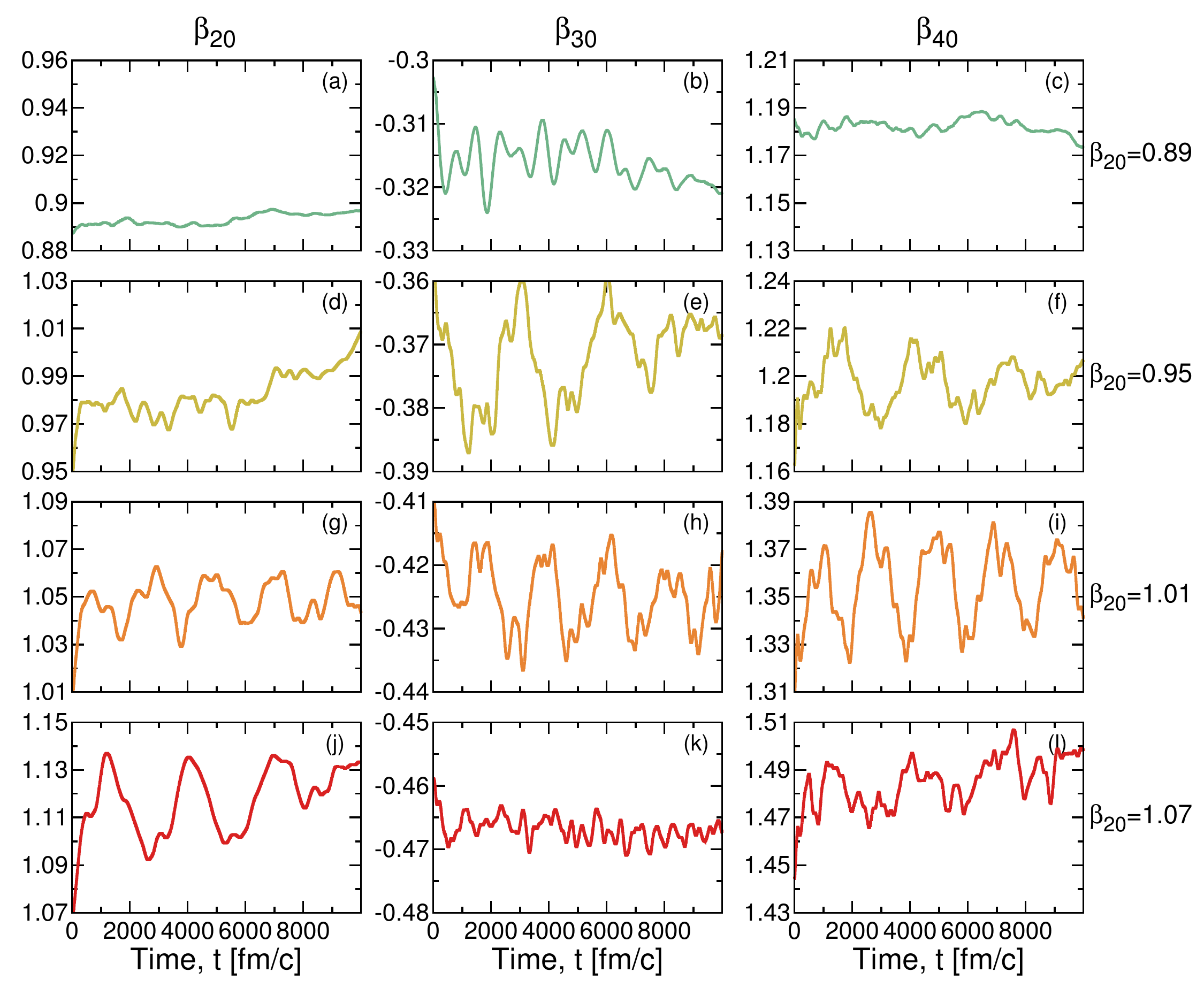} 
\caption{(Color online) Time evolution of quadrupole (left columm), octupole (centre columm) and hexadecapole (right columm) multipole parameters for initial states which are solutions to CHF calculations, with initial quadrupole deformation labelled on the right-hand side. All of the initial states are deformed beyond the peak of the static fission barrier, $\beta_{20}>0.86$. }
\label{res-zoom-above}
\end{center}
\end{figure*}

Beyond the peak of the static fission barrier, the time evolution of several increasingly deformed initial states fail to display fission within $9000$ fm/$c$. The evolution of the multipole deformation parameters for these states is presented in Fig.\@ \ref{res-zoom-above}.

Qualitatively, one observes dramatically different behaviour in the time evolution of the multipole deformations compared to a giant-resonance-like behaviour below the second barrier. For most cases, the elongation is seen to increase rapidly during the first $300$ to $500$ fm/$c$, indicated by an increase of $\beta_{20}$. The most extreme case is seen in the bottom left panel of the figure, where the quadrupole deformation quickly increases from $\beta_{20}\approx1.07$ to $\beta_{20}\approx1.11$. 

Beyond the initial increase in elongation, Fig.\@ \ref{res-zoom-above} displays slow, large amplitude oscillations setting in. Compared to the states below the static fission barrier, these oscillations are substantially slower. They therefore correspond to lower energy modes in the power spectrum. For the initial configurations with $\beta_{20}=0.89$ and $0.95$ (panels (a) to (f) in Fig.\@ \ref{res-zoom-above}), the behaviour of the quadrupole deformation is more complex than the other two cases ($\beta_{20}=1.01$ and $1.07$). The evolution of the quadrupole deformation for these cases ($\beta_{20}=0.89$ and $0.95$) shows a region of rapid increase, then an oscillation about a plateau, then another rapid increase followed by another plateau. 

An octupole deformation (depicted in the central column of Fig.\@ \ref{res-zoom-above}) is observable in all cases. This is unsurprising in itself, as the initial configurations are significantly octupole deformed. However, an interesting feature is noticeable for the evolution of the states with initial deformation $\beta_{20}=0.95$ and $1.01$. The changes in octupole deformation are roughly in phase with either the evolution of the hexadecapole parameter or both the quadrupole and the hexadecapole parameters. This feature, in addition to the other differences observed, suggests that the mechanism driving the dynamics  of the multipole deformations above and below the fission barrier is different.

The slow, large-amplitude oscillatory behaviour of the multipole deformation parameters (second row from bottom of Fig.\@ \ref{res-zoom-above}) suggests that, owing to the Coulomb repulsion between the upper and lower lobes, the nucleus is attempting to fission. This is in line with the macroscopic model of Bohr and Wheeler \cite{Boh39}, where the effect of the charge on an incompressible liquid drop is a crucial ingredient to describe the fissioning process. Within macroscopic liquid drop models, the surface term competes with the repulsive Coulomb force to inhibit fission; it costs energy to form an increasingly deformed shape. The TDHF calculations present a similar behaviour, but the mechanism arises microscopically.  
Further calculations, not shown here for brevity, confirm the relevant role of Coulomb repulsion in the dynamics of these configurations \cite{Goddard14}.

Overall, there is significant evidence concerning the mechanism responsible for the slow, large-amplitude oscillations of the nuclear shape observed on non-fissioning configurations beyond the static fission barrier. The oscillations are driven by a competition between the Coulomb force, trying to cause fission, and the attractive nuclear potential terms in the energy functional, countering this effect. When the states are evolved in time, they explore significant collective motion in an attempt to find a pathway towards fission. It can only be speculated as to whether, with a long enough time evolution, the states would eventually fission. Panel (j) of Fig.\@ \ref{res-zoom-above} (evolution of a state from $\beta_{20}$=1.07) shows that the quadrupole deformation oscillates around a gradually increasing average. This suggests a final state that could eventually fission. The timescale for the quadrupole increase is very slow, however. A TDHF calculation that explores the time evolution beyond $t \approx 10,000$ fm/$c$ may begin to encounter numerical instabilities. 

\subsection{Intersection of the One and Two-Fragment Fission Pathways}
\label{1vs2int}

\begin{figure}[tbh]
\begin{center} 
\includegraphics[width=\linewidth]{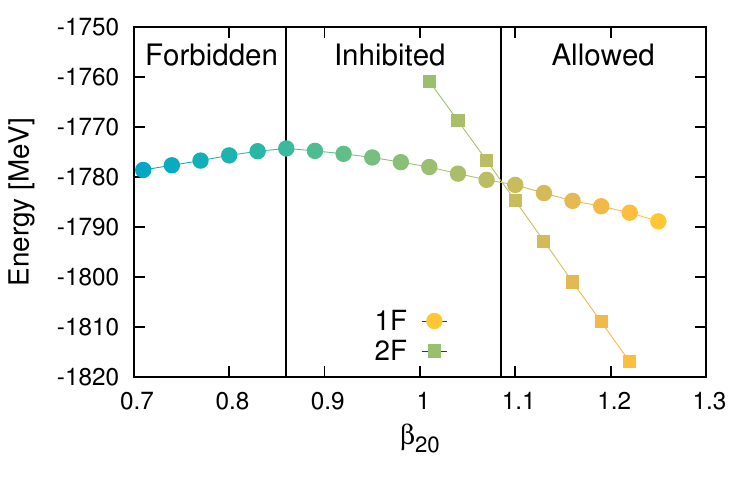} 
\caption{(Color online) Static one- and two-fragment fission pathways (solid circles and squares, respectively). For TDHF calculations, three regions may be defined when following the one-fragment pathway. In a first region, fission is forbidden within TDHF time scales. Beyond the maximum in the PES at $\beta_{20}=0.86$, the time scale for fission is inhibiting. Finally, beyond the crossing of two pathways at $\beta_{20}=1.085$, fission is allowed. See text for more details.}
\label{intersect_path}
\end{center}
\end{figure} 

One striking characteristic of the static one- and two-fragment pathways is the intersection point that separates the initial states which fission upon time evolution from those which do not. Figure \ref{intersect_path} zooms into the area of the PES which is relevant for these differences. We divide the PES into three regimes. For configurations starting with a deformation below the static fission barrier ($\beta_{20} <0.86$), tunnelling is required to reach a fissioned state. This is forbidden in TDHF calculations, in which collective coordinates behave semiclassically. We therefore define a forbidden region below the barrier. Beyond the barrier, in contrast, there is a region where fission is inhibited. It will shortly be demonstrated {in Sec. \ref{fissin}} that, beyond the intersection of the one and two-fragment pathways, fission is allowed within the considered time scales of the TDHF calculations. The line between $\beta_{20}=1.07$ and $1.10$ thus separates the inhibited and allowed regions for fission, but the deformation itself does not correspond to a threshold in the dynamic calculations. The bottom row of Fig.\@ \ref{res-zoom-above} indicates, for example, that the state with an initial deformation just below the separating line can evolve dynamically to a state with deformations beyond this very same line, but without fissioning.

\begin{figure*}[tbh]
\begin{center} 
\includegraphics[width=0.6\linewidth]{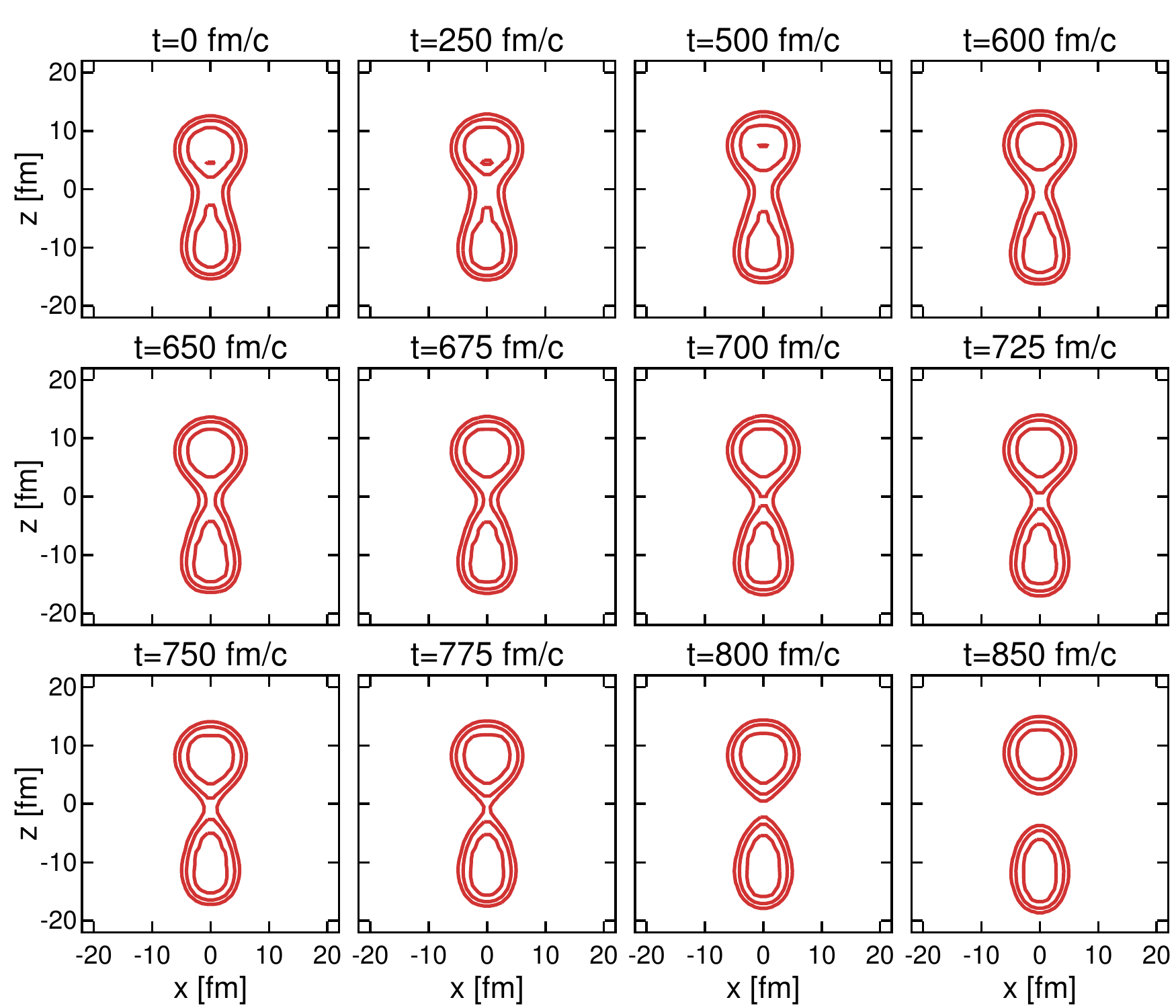} 
\caption{(Color online) Slices of the total particle density for various times, starting from the static case with $\beta_{20} = 1.19$. It takes between 775 and 800~fm/$c$ for scission (as defined in the text) to occur. The isolines are separated by 0.05 particles/fm$^{3}$. }
\label{fiss-relasing}
\end{center}
\end{figure*}

An intuitive explanation may be given for the significance of the intersection point of the pathways on the PES with regard to fission occurring upon time evolution. In TDHF, the total energy is conserved. It is because of the inclusion of excitations (internal and translational) that nuclear configurations may change upon time evolution. For the states which undergo fission (allowed region in Fig.~\ref{intersect_path}, $\beta_{20}>1.085$), for a given value of $\beta_{20}$, the two-fragment state is \textit{more} bound than the one-fragment state. In consequence, the one-fragment state can evolve into a two-fragment configuration at a constant $\beta_{20}$ by releasing nuclear potential energy into excitation energy. Of course, the picture is not really that simple, because the significance of the static PES becomes less clear in the dynamic case. In general, configurations which do not correspond to the static fission pathways will be explored. Further, a slight change of configuration will be required to move from the static one-fragment state to a fissioned configuration. In other words, the exact configurations on the two-fragment pathway cannot be reached dynamically from the one-fragment pathway, but an excited two-fragment configuration of a similar deformation can. The reasoning presented is that, as the static two-fragment state is more bound than the corresponding one-fragment state, the optimum TDHF trajectory is to evolve the one-fragment static state towards an excited fissioned configuration by undergoing only a modest rearrangement of the nuclear shape.

In contrast, in the inhibited region of Fig.\@ \ref{intersect_path} ($0.86 \le\beta_{20} \le 1.085$), for a given $\beta_{20}$ in the one-fragment pathway, the two-fragment state with the same $\beta_{20}$ is \textit{less} bound. Owing to energy conservation, the one-fragment state cannot move to the two-fragment state at the same $\beta_{20}$. The only way to reach a two-fragment solution of equal binding energy (or an excited configuration with greater binding energy) is through a significant change in deformation and rearrangement of the nuclear state, which accounts for the inhibiting time scale for fission to occur.  It may be that the presented calculations lack the degrees of freedom necessary to allow a fissioning path to be found in this window at all, and that a method beyond basic TDHF, including either collisions \cite{Wong1979} or dynamical pairing effects \cite{Scamps2015}, is needed to reach the fissioned configuration. Further investigation of the link between static and dynamic configurations applying density-constrained TDHF \cite{Cus85,Uma10}, would certainly be of interest. This method allows the dynamic configurations to be ``frozen", removing internal and collective excitations, thus bridging between static and dynamic configurations.

In our frozen BCS approach, however, the lack of dynamical single-particle occupation effects could be particularly important in this region of the PES. Relatively compact initial states can have substantially different single-particle structure as compared to the real fissioning fragments. The dynamical final state that we generate is therefore only an approximation to the real one, which will be captured better in simulations including time-dependent superfluidity \cite{Scamps2015}. While the dynamics in the interface between the inhibited and allowed regions can be considered somewhat artificial, we expect these effects to be less important for initial states well into the allowed regime, $\beta_{20}>1.15$. There, the structure of the pre-fission system is already reminiscent of a two-fragment state, and dynamical rearrangement of single-particle orbits is likely to be less important. 

 \subsection{Fissioning States}
\label{fissin}

\begin{figure*}[tbh]
\begin{center} 
\includegraphics[width=0.7\linewidth]{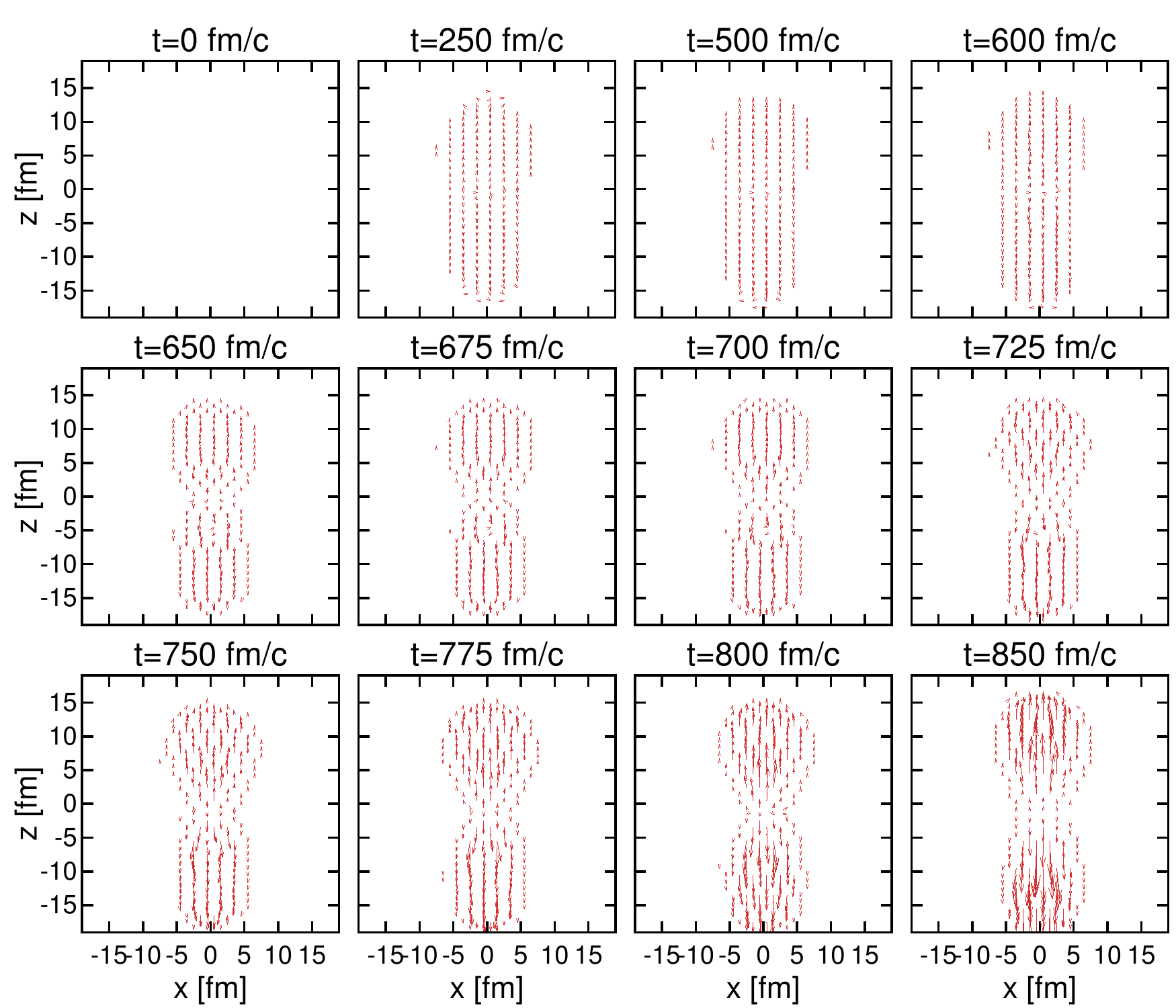} 
\caption{
(Color online) Current vectors corresponding to the slices of the particle density presented in Fig.\@ \ref{fiss-relasing}. The vectors have been normalised to the same scale in each panel, to a visually instructive length. Note that the dimensions of the grid exceed what is presented in these panels. 
\label{fiss-relasing-current}}
\end{center}
\end{figure*} 

For the static states with a quadrupole deformation at and beyond the threshold of $\beta_{20}=1.10$, binary fission is seen to occur as the wave functions are evolved in time. The calculations to obtain the data for this Section were performed in a larger grid of size $48\times48\times160$ points, corresponding to $-79.5$ to $79.5$ fm in the $z$ direction, and $-23.5$ to $23.5$ fm in the $x$ and $y$ directions. The calculations were set to end once the separation of the center of mass of the two fragments exceeded $100$ fm. This cutoff avoids spurious effects associated with the fragments approaching the grid boundaries.

Figure \ref{fiss-relasing} shows the typical time evolution of the particle density for the fissioning case by presenting 2D slices of the 3D density at various times for the state with initial deformation $\beta_{20}=1.19$. It is difficult to establish exactly the scission point in a calculation involving quantum mechanical wave functions and densities. We take an operational approach and define ``scission" as the point when the minimum density between the fragments along the principal axis of the system is less than $0.05$ particles/fm$^{3}$. As we shall see in the following, this also corresponds to the point where a sizable collective energy develops as the fission products begin spatially separating. For the case where the initial quadrupole deformation is $\beta_{20}=1.19$ (presented in Fig.\@ \ref{fiss-relasing}), it takes between $775$ and $800$ fm/$c$ for the density between the two fragments to drop below this threshold. Figure \ref{fiss-relasing-current} displays sample current vectors corresponding to the particle density slices of Fig.\@ \ref{fiss-relasing}. The current vectors display the system smoothly transitioning into a two-fragment configuration. There is no dramatic rearrangement of the density during time evolution. Throughout the calculation, the currents in the two preformed fragments are clearly distinguishable, and do not interact with one another. The central region has negligible current, and the two lobes stretch against each other. The magnitude of the current vectors in Fig.\@ \ref{fiss-relasing-current} gradually increases as the fission occurs. Beyond the point of scission, they increase rapidly as the fragments accelerate away from one another. 

\begin{figure}[tbh]
\begin{center} 
\includegraphics[width=0.7\linewidth]{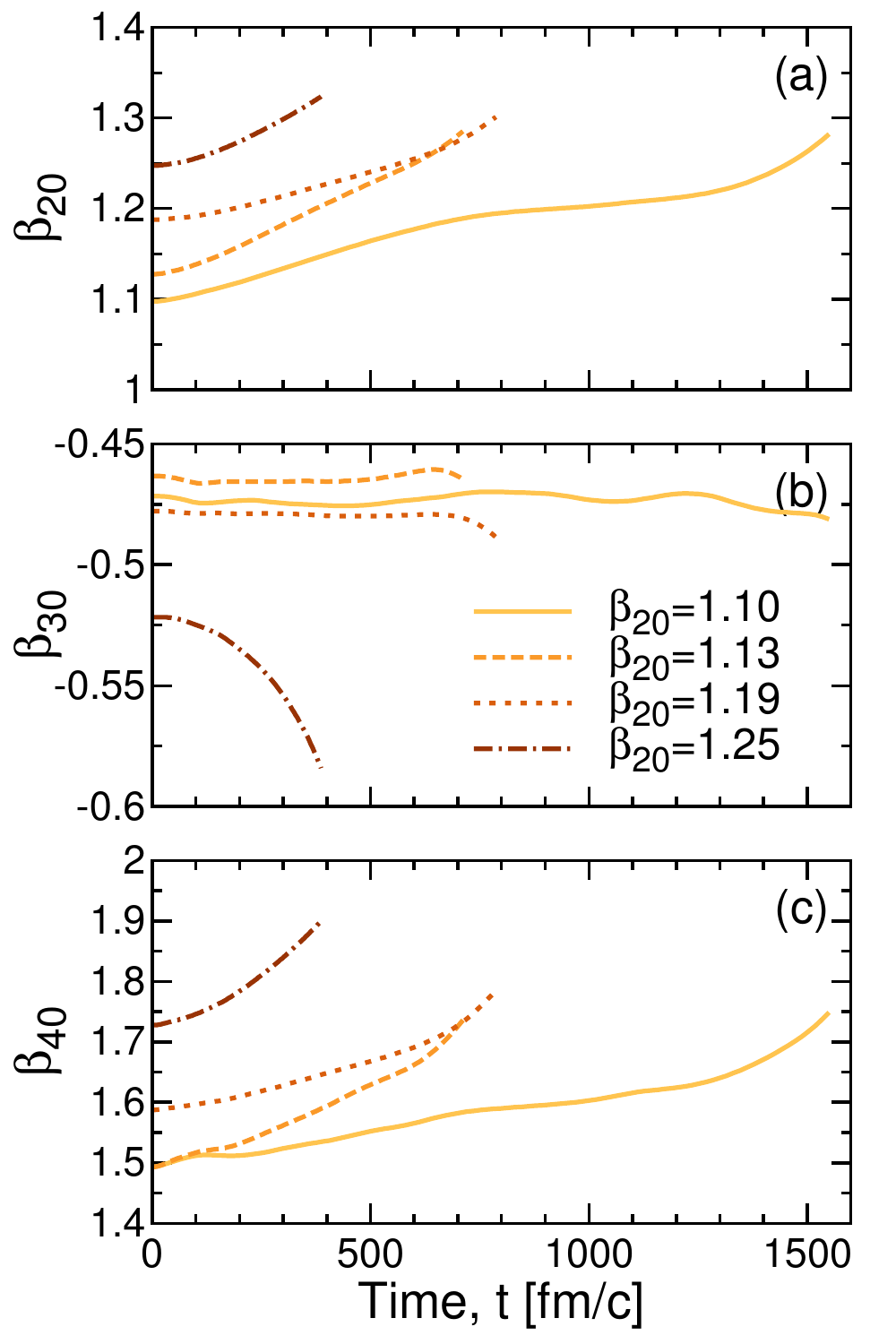} 
\caption{(Color online) Time evolution of (a) quadrupole, (b) octupole and (c) hexadecapole deformation parameters for various static configurations observed to fission upon time evolution. The evolution is stopped at the scission point, as defined in the text.}
\label{fiss-mpole}
\end{center}
\end{figure}

States with increasingly large initial deformations fission in a qualitatively similar way. We now proceed to compare their fission outcomes in terms of macroscopic observables. The states with static deformation $\beta_{20}=1.10, 1.13, 1.19$ and $1.25$ were evolved in time to investigate the fission of the different initial configurations. The time evolution of the multipole moments for these states are shown in Fig.\@ \ref{fiss-mpole}. Different nuclear shapes are explored as the nucleus evolves from the various static states. Other than the case with initial $\beta_{20}=1.25$, we find that as $\beta_{20}$ and $\beta_{40}$ increase, the octupole deformation, as reflected in $\beta_{30}$, remains virtually constant. Here and in the following discussion,  we sharply cut off the time evolution at the point of scission. An analysis of the post-scission fragments is presented in Sec.\@ \ref{sec:fragments}. 

The chosen dynamic pathway towards fission, depending on how the particles rearrange during the time evolution, may have significant consequences upon the properties of the post-fission system. This will produce a range of fission fragments dependent on the initial configuration which is time evolved. Once again, this differs from the static case, where CHF calculations following the one-fragment fission pathway will only produce one resulting fissioned configuration. The distribution of fission products obtained with TDHF is in line with experimental investigations (see Sec.\@ \ref{sec:fragments}).

The time scale required for the initial configuration to fission varies. The least elongated case, with $\beta_{20}=1.10$ takes $t_\text{scission} \approx 1250$ fm/$c$ for scission to occur. Figure \ref{fiss-mpole} shows the quadrupole deformation increasing rapidly for approximately $600$ fm/$c$ [panel (a), solid line]. Between $600$ and $1200$ fm/$c$, the rate of increase in quadrupole deformation reduces as the nucleons rearrange out of the neck into the upper and lower fragments. Small oscillations in the octupole deformation can be seen as the system transitions into the preferred configuration. Beyond $1200$ fm/$c$, the neck rapidly vanishes as the fragments take form and begin to separate, resulting in an acceleration in the increase of the quadrupole and hexadecapole parameters.

For more deformed initial states, the time taken to fission is significantly shorter. This can be explained as the initial configuration has fewer particles in the neck region. Upon time evolution, less rearrangement is required for the two fragments to take form, and the Coulomb interaction rapidly drives the configuration to fission. The most extreme case investigated here is that with initial deformation $\beta_{20}=1.25$. Panel (k) of Fig.\@ \ref{pu-esurf-pretty} shows the initial density for this state. Two fragments are already taking form, connected only by a thin elongated neck. This narrow structure rapidly dissipates into the top and bottom fragments upon time evolution, and within $t_\text{scission} \approx 400$ fm/$c$ the system fissions.

\begin{figure}[h!]
\begin{center} 
\includegraphics[width=\linewidth]{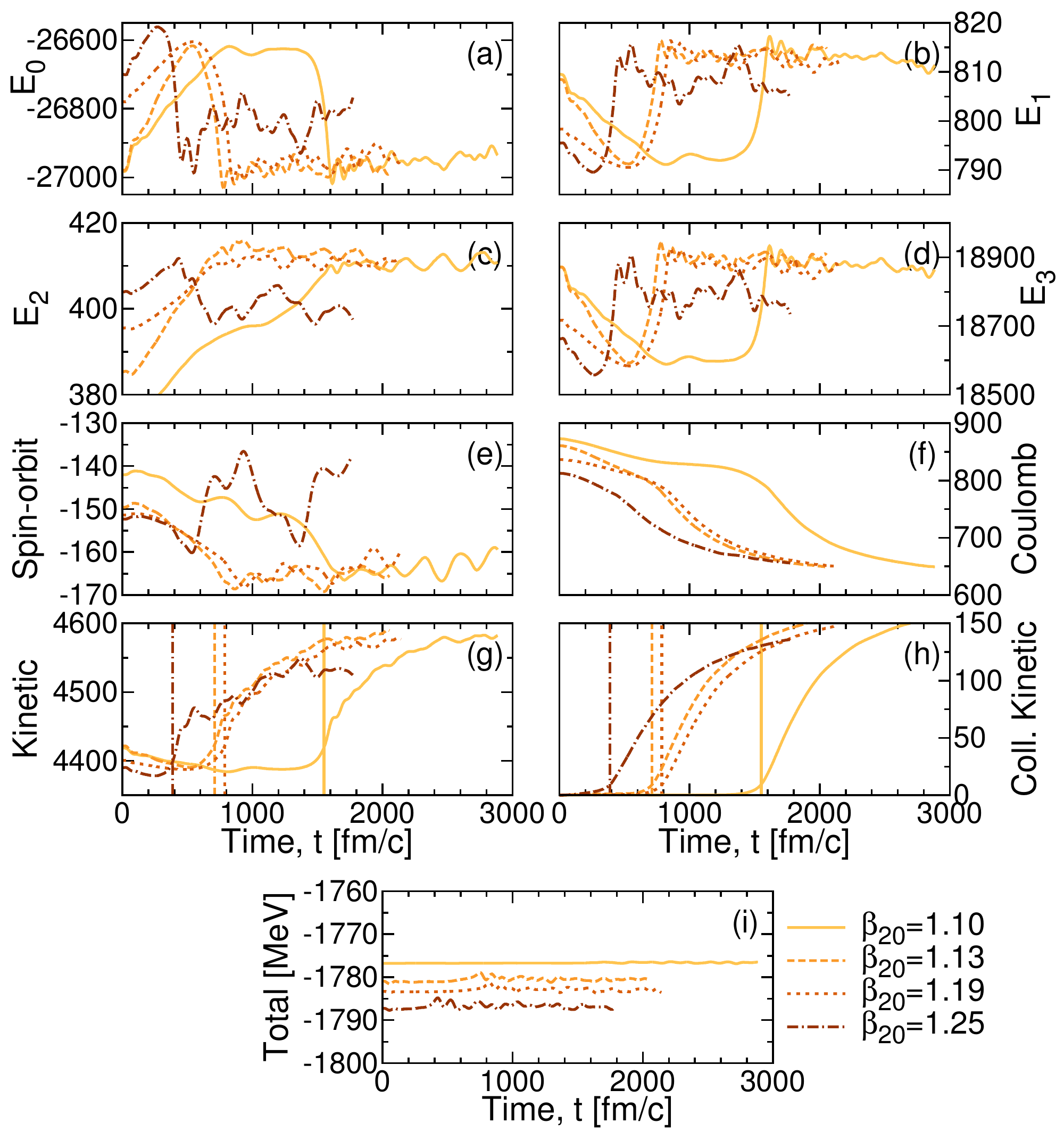} 
\caption{
(Color online) Evolution of the decomposed energy density functional for the fissioning systems. The calculations are terminated when the fragments are separated by 100 fm. For reference, the vertical lines in the panels corresponding to kinetic energies show the corresponding scission points. The total energy is conserved within fluctuations no greater than 4 MeV.}
\label{fiss-edf}
\end{center}
\end{figure} 

The evolution of the decomposed contributions to the energy density functional for the fissioning cases are presented in Fig.\@ \ref{fiss-edf}. The decomposed energy density functional for the entire two-fragment system is shown. The total energy of the two separate fission fragments will be analysed separately in Sec.\@ \ref{sec:fragments}. Figure \ref{fiss-edf} displays the evolution of the energy density functional up to and {\it beyond} the point of scission. The total energy, shown on the panel (i), is, in principle, conserved within the TDHF calculations. For these fission calculations, the fluctuations during the time evolution are less than 4 MeV.

The dynamic calculations allow for both translational motion and internal excitations. In the fissioning case, the nuclear binding energy is expected to be transformed mainly into the translational kinetic energy of the fragments. The nuclear collective kinetic energy is conventionally defined within TDHF as,
\begin{equation}
E_{\text{coll. kin.}} = \frac{\hbar^2}{2m} \int d \bm r  \, \frac{\bm j(\bm r)^2}{\rho(\bm r)} \, ,
\label{ecooll}
\end{equation}
where $\rho(\bm r)$ is the particle density and $\bm j(\bm r)$ is the current density \cite{Mar13}. This collective kinetic energy contains contributions from internal excitations of the nucleus, such as resonances, as well as the translational kinetic energy of the post-fission fragments. It is presented in panel (h) of Fig.\@ \ref{fiss-edf}. The total kinetic energy is shown separately, in panel (g). It is difficult to untangle the collective excitation energy attributed to the internal excitation of the fission fragments, from that attributed to translational motion. In fission reactions, the energy release is typically attributed to be $\approx 80 \, \%$ in the form of translational energy, and the other $\approx 20 \, \%$ is released in the form of $\gamma$ rays, prompt neutron emission, and radioactive decays of the fragments \cite{Kra88}.  In our TDHF calculations, only part of these effects can be described. The degrees of freedom to allow fission fragments in hot resonance states to decay by particle emission, for instance, are included.  We demonstrate that the excitation energy of the fissioned system in the TDHF simulations is dominated by the translational kinetic energy, with a small contribution from internal collective excitation of the fragments. This will be discussed in Sec.~\ref{coll-ex-frag}. 

During the time evolution, the individual components of the energy functional may be examined separately. The physical interpretation of the evolution of each term of the integrated energy functional shown in Fig.\@ \ref{fiss-edf} may not necessarily be simple. It is useful to identify which densities contribute to the separate terms to qualitatively explain the behaviour of the energy functional: 

\begin{itemize}
\item \underline{$E_0$ and $E_3$ terms}: These terms, proportional to the Skyrme parameters $t_0$ and $t_3$, are the central terms of the functional. They are attractive and repulsive,1 respectively [see panels (a) and (d)], providing a similar functional form with an expected cancelation between them.  They depend upon the particle density. As the nucleus approaches scission, the $E_0$ term is reduced in strength, thanks to the small, unfavourable density that briefly exists in the neck region.  There is then a sudden increase in biding as separate, more stable fragments are formed.  After this, the energy contributions oscillate with a smaller magnitude than the changes during the fission process, corresponding to the excited collective motion of the fragments
\item\underline{$E_1$ term}: The $E_1$, shown in panel (b), term contains contributions from the kinetic, particle, and current densities. The time evolution of this term is qualitatively similar to the $E_0$ and $E_3$ contributions, with the same sign as the $E_3$ term but an absolute smaller scale. This suggests the that density that governs the $E_0$ and $E_3$ terms, the particle density,  is also the most relevant contribution driving the $E_1$ term.
\item\underline{$E_2$ term}: The $E_2$ contribution in panel (c) contains the Laplacian of the particle density, and is commonly associated with a surface term. As the particles rearrange into the two fission fragments, this term increases in magnitude - i.e. gives an overall more repulsive contribution to the entire system. This can be explained by the two-fragment system having a combined surface fraction which is greater than that of the initial configuration. The gain in energy for this term up to the point of scission is dependent upon the deformation of the initial configuration. It can increase by as much as $45$ MeV for the static configuration with $\beta_{20}=1.10$. 
\item\underline{Coulomb term}: The Coulomb energy depicted in panel (f) is determined from the distribution of the charged protons. The magnitude of the repulsive Coulomb term slowly decreases as the nucleus elongates. The overall reduction is of the order of $200$ MeV as the system evolves. At the point of scission, the rate at which the term reduces rapidly accelerates, as the two charged fragments separate from one another in co-ordinate space. At infinite fragment separation, the Coulomb term will reduce to the contributions of the Coulomb energy for each nucleus, without further interactions. 
\item\underline{Kinetic and Collective Kinetic terms}: The kinetic energy can be determined from integrating the kinetic density. As mentioned above, the contribution to this energy from collective motion (assumed to be predominantly translational beyond scission, rather than internal collective excitation) can be decomposed according to Eq.\@ (\ref{ecooll}) and we show it in panel (h). The collective energy is initially small, as it is only associated with the internal currents as the nucleus slowly rearranges into a fissioned configuration. The state with initial deformation $\beta_{20}=1.10$ shows the most gradual transition to fission. An initial rapid increase and a saturation in the collective energy is seen before scission, which corresponds to the (previously discussed) rapid initial elongation, then extended rearrangement phase as the configuration evolves (see Fig.\@ \ref{fiss-mpole}). In contrast, the state with $\beta_{20}=1.25$ is already close to the point of scission, so that the Coulomb interaction between the two lobes rapidly drives the configuration to the scission point (within the first few hundred fm/$c$), where translational motion rapidly accelerates once the neck ruptures. This shorter timescale could explain the more extreme behaviour observed in the evolution of the other terms in the energy functional as the particles in the neck have less time to rearrange into the two fragments. At the point of scission, the collective kinetic energy rapidly increases at a rate similar to that of the reduction in the Coulomb energy. The threshold collective kinetic energy associated with the scission point is between 6 and 8 MeV in all the cases presented. The vertical lines corresponding to the scission points are displayed in panels (g) and (h) of Fig.\@ \ref{fiss-edf}. The definition adopted for scission \footnote{Point at which neck density is below 0.05 particles/fm$^3$ along the principal axis of the system.} is justified by considering the rapid increase of collective kinetic energy at this point. The gain in the total kinetic energy beyond the scission point can be attributed to the gain in collective energy, and is of the order of 150 MeV in the time considered. This relates to the loss in Coulomb energy \textit{beyond} the point of scission, as would be expected.
\item\underline{Spin-Orbit term}: The spin-orbit potential is important for the single-particle structure.  Spin-orbit partner levels combine, when fully occupied, to give a total zero energy contribution, so the contributions reflect the changing of the single particle levels and the change of meaning of spin-orbit partners in the parent nucleus to the daughter nuclei. One therefore expects this contribution to be relevant in determining the structural details of the final fission fragments. The final approximately constant values observed in panel (e) following scission correspond to the sum of the two independent spin-orbit terms of the separate fragments. The notably different behaviour of the term for the state with initial deformation $\beta_{20}=1.25$, compared to the others, suggests that different shell effects are acting. Indeed, the masses of the fission products are significantly different for this case (see Sec.\@ \ref{DIFmassdist}). The spin-orbit term also has significant contributions from time-odd densities, so that the evolution can explore configurations which may not be accessible on the static PES. Overall, the term varies by less than 30 MeV during time evolution. 
\end{itemize}

In experimental studies of fission, it is customary to measure the kinetic energy of the fission fragments. We can find an analogous observable within TDHF by making use of the collective kinetic energy, defined by Eq.\@ (\ref{ecooll}), and assuming that the translational kinetic energy dominates this term. Referring to Fig.\@ \ref{fiss-edf}, the collective kinetic energy and Coulomb energy are both expected to plateau as the separation of the two fragments becomes large. Unfortunately, the Coulomb force is long-ranged. Because increasing the dimensions of the numerical grid is extremely computationally expensive, we take an alternative perspective. We extrapolate the collective kinetic energies to large times to estimate the asymptotic value of the collective energy as the separation $r$ tends to $\infty$.

\begin{figure}[tbh]
\begin{center} 
\includegraphics[width=\linewidth]{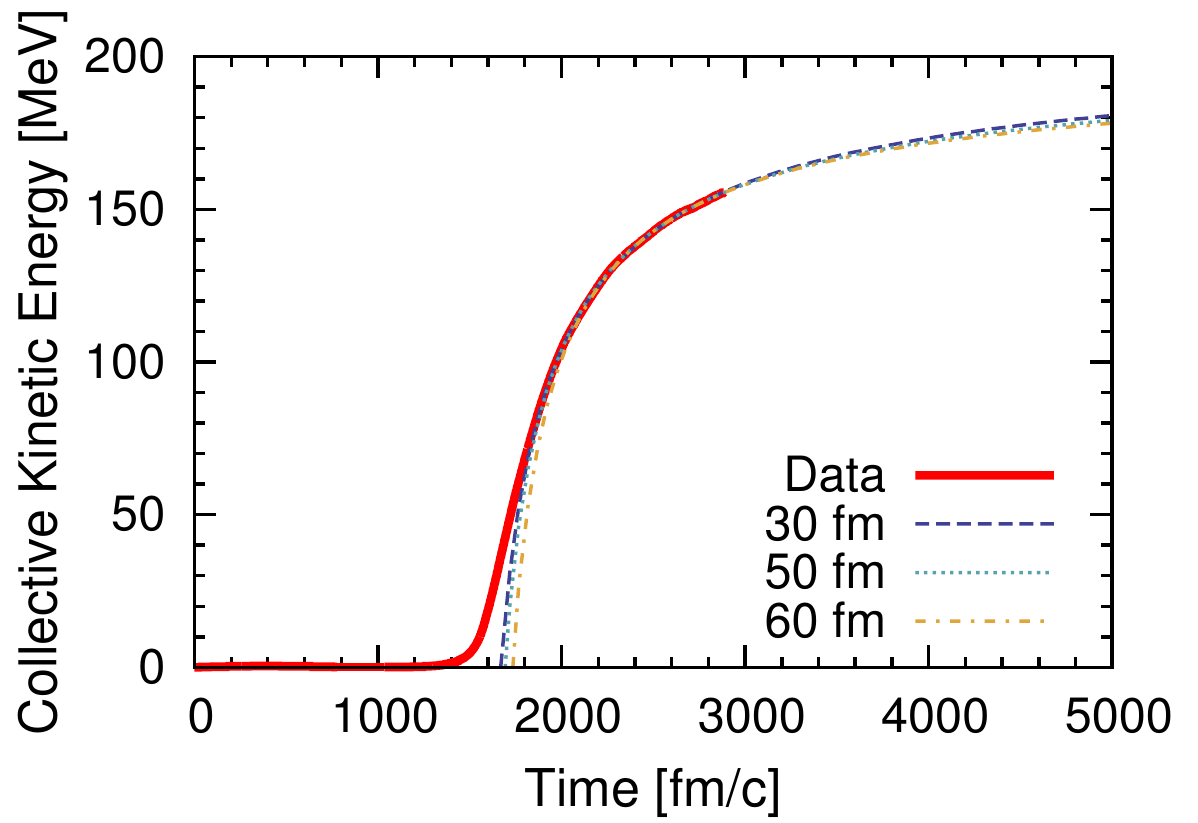} 
\caption{(Color online) Fits to the obtained collective kinetic energy for the initial state with deformation $\beta_{20}=1.10$. Fits are performed over three different ranges: from the point where the separation of the fragments exceeds $30$, $50$ and $60$ fm, respectively. }
\label{fiss-kin}
\end{center}
\end{figure}

A simple approximation to the time-dependent behaviour can be made from classical mechanics. Let us assume two pointlike fragments with charges $Z_u$ and $Z_l$, and masses $M_u$ and $M_l$. If the two fragments fission from a ground state because of the Coulomb force, and convert all this energy into translational kinetic energy, energy conservation implies: 
\begin{equation}
\label{econs}
\frac{1}{2}M_uv_u^2 + \frac{1}{2}M_lv_l^2 = \kappa \frac{Z_uZ_l}{r} \, .
\end{equation}
Here, $M_i$, $v_i$ and $Z_i$ are the mass, velocity and charge of each fragment ($i=u$ for upper and $i=l$ for lower fragment). The constant $\kappa$ is the Coulomb constant. As momentum must be conserved,
\begin{equation}
M_uv_u+M_lv_l = 0 \, ,
\end{equation}
Eq.\@ (\ref{econs}) may be rewritten, substituting for $v_u$
\begin{equation}
v_l^2\left(\frac{M_l^2}{M_u}+M_l\right) = 2\kappa \frac{Z_uZ_l}{r} \, .
\end{equation}
For a given fissioned system, $M_u,M_l,Z_u$ and $Z_l$ are constant. A differential equation for $\frac{d r}{dt}(=v_l)$ can be formed 
\begin{equation}
\frac{d r}{dt} = \sqrt{\frac{\Theta}{r}} \, ,
\end{equation}
where all the constants are combined into $\Theta$. Performing the integration
\begin{equation}
\int_{r_0}^r r^{1/2} d r = \int _{t_0}^t \sqrt{\Theta} d t
\end{equation}
allows the solution
\begin{equation}
r^{3 / 2} = r_0 ^{3/2}  + \frac{3}{2} \sqrt{\Theta} (t-t_0) 
\end{equation}
to be written. According to this approximation, the distance between the two fission fragments, $r$, is approximately proportional to $t^{2/3}$. By assuming that the loss in Coulomb energy is equal to the gain in collective kinetic energy (that is , $E_{\text{Coul}}=E_{\text{coll. kin.}}$), a fit of the form 
\begin{equation}
f(t) = a + \frac{b}{(t-c)^{3/2}}
\label{fitfunc}
\end{equation}
can be performed to interpolate the collective kinetic energy to asymptotically large values of $t$. Figure \ref{fiss-kin} shows a sample interpolation of the collective kinetic energy assuming the above form for the case of initial deformation $\beta_{20}=1.10$. The fit is performed over three time ranges: once the centers of mass are separated beyond $30$, $50$ and $60$ fm, respectively. As the separation tends to $\infty$, the fit parameter $a$ can be interpreted as the final collective kinetic energy. Table \ref{frag-ke-tab} contains the values obtained for each of the fissioning cases with different distance fits. As a crude method to represent the uncertainty in the final collective energy value, we present the mean and the standard deviation of the three interpolation fits. 

The values shown in Table \ref{frag-ke-tab} demonstrate that the resulting collective kinetic energy varies by about $10 \%$ depending upon the region of the data the fit is performed to.  In this very crude model, fragment deformation, particle emission (discussed in the next section) or tidal effects associated with the extended nature of the nuclei are not accounted for. This suggests that the results obtained from the interpolation method should serve only as illustrative values. We note that an alternative method, based on computing the collective kinetic energy from the center-of-mass momenta of each fragment, provides quantitatively similar results.

\begin{table}[tbh]
\caption{Interpolated total kinetic energy corresponding to different initial configurations. The fit of Eq.\@ (\ref{fitfunc}) has been performed once the fragment separation exceeds 30 fm, 50 fm and 60 fm.}
\centering
\begin{tabular}{|c|c|c|c|c|}
\hline 
 Static & Coll. KE  & Coll. KE & Coll. KE & Mean  \\
Deformation  &(30 fm fit )&(50 fm fit )&(60 fm fit ) &$\pm$ St. Dev. \\
 $\beta_{20}$ &[MeV]&[MeV]&[MeV]  &[MeV]\\
\hline\hline
1.10&210.3&206.5 & 203.4& 206(4) \\
1.13&210.8&200.0&193.7& 202(8)  \\
1.19&205.8&196.8&191.3&  198(8)\\
1.25&193.4 &180.8&176.3& 183(9) \\
\hline 
\end{tabular}
\label{frag-ke-tab}
\end{table}

In comparison to the experimentally measured kinetic energy of the fissioning systems displayed in Table \ref{frag-ke-exp-tab1}, the theoretical values presented in Table \ref{frag-ke-tab} are of a similar order of magnitude. The crude theoretical estimates are about $5$ to $25$ MeV larger in all cases. We note, however, that the experimental values correspond to an average kinetic energy. We only have access to a single fissioning event per static state, and a larger sample of theoretical results would be required to enable a quantitative comparison. Further discussion of methods to deduce the energy released by the fission reaction within TDHF is presented in Sec.\@ \ref{fragesec}.

\begin{table}[h!]
\caption{Measured total kinetic energies from experiments on spontaneous fission, thermal neutron-induced fission and photo-fission. The measurements correspond to the pre-neutron emission fragment energies.} 
\centering
\begin{tabular}{|c|c|c|}
\hline 
 Method& Kinetic energy [MeV] & Reference \\
\hline\hline 
$^{240}$Pu(s.f) & 178.85$\pm$0.30 & \cite{Thi81} \\
$^{240}$Pu(s.f) & 179.00$\pm$0.08 & \cite{Wag84} \\
$^{239}$Pu($n_{th},f$)&177.69&\cite{Thi81} \\
$^{239}$Pu($n_{th},f$)&177.65$\pm 0.01$&\cite{Wag84}  \\
$^{240}$Pu($\gamma,f)$ (12 MeV)&176.39$\pm$0.24&\cite{Thi81} \\
$^{240}$Pu($\gamma,f)$ (15 MeV)&175.80$\pm$0.24&\cite{Thi81} \\
$^{240}$Pu($\gamma,f)$ (20 MeV)&175.15$\pm$0.24&\cite{Thi81} \\
$^{240}$Pu($\gamma,f)$ (30 MeV)&174.98$\pm$0.31&\cite{Thi81} \\
\hline 
\end{tabular}
\label{frag-ke-exp-tab1}
\end{table}

\section{Fragment Analysis}
\label{sec:fragments}

Beyond the point of scission, we consider a two-fragment system. The published distribution of {\sc sky3d} has some capacity to analyse two-fragment dynamics \cite{Mar13}, and a version has been modified further to investigate the fissioning system and extract some useful observables \cite{Goddard14}. 

\subsection{Mass Distributions}
\label{DIFmassdist}

\begin{table*}[tbh]
\caption{Fission fragments obtained from evolving initial static configurations from the one-fragment fission pathway. The uncertainties in the particle numbers are a conservative estimate related to the fluctuation in the particle number in the region of the grid corresponding to the separate fragments throughout time evolution. The result from the static two-fragment pathway for $\beta_{20}=1.19$ is included for comparison. } 
\centering 
\begin{tabular}{|c|c|c|c|c|}
\hline 
 Static &Heavy Fragment&Light Fragment& Heavy Frag. & Light Frag.   \\
 Deformation $\beta_{20}$&$A,Z$&$A,Z$&(Integer)&(Integer) \\
\hline\hline
1.10&136.33(5)  ,  52.78(5)&103.67(5)  ,  41.23(5)&$^{136}_{53}$I&$^{104}_{41}$Nb \\
1.13&135.02(5)  ,  52.23(5)&104.98(5)  ,  41.77(5)&$^{135}_{52}$Te&$^{105}_{42}$Mo\\
1.19&136.13(5)  ,  52.70(5)&103.87(5)  ,  41.30(5)&$^{136}_{53}$I&$^{104}_{41}$Nb\\
1.25&143.70(5)  ,  55.65(5)&96.30(5)  ,  38.35(5)&$^{144}_{55}$Cs&$^{96}_{38}$Sr\\
1.19$^\text{(2f)}$&132.81  ,  50.84&107.04  ,  43.13&$^{133}_{51}$Sb&$^{107}_{43}$Tc\\
\hline 
\end{tabular}
\label{frag-mass}
\end{table*}

As the post-fission fragments are excited, they may decay by particle emission. TDHF displays this decay by the spreading of the single particle wave functions from the region of central density, corresponding to the nucleus. When masking the region around the nucleus, this decay results in a reduction in the integrated particle density over time. For the cases of DIF examined in this paper, this decay is of the order of $0.1-0.2$ particles during the whole postscission time evolution. 

To compare to experimental studies, we identify the number of particles in each fragment prior to any particle emission. This is done by integrating the total density in each half of the numerical grid separated by the dividing plane immediately after the scission point. The integral is performed without any masking. An uncertainty in the particle number of the fragments may be associated with the fluctuation of this measurement throughout time evolution, which is less than $0.05$ particles for all the considered cases. We are thus confident that the fission fragments that we produce have a good average mass number. 

\begin{table*}[tbh]
\caption{
Experimentally measured average masses following the fission $^{240}$Pu. The measurements for neutron-induced fission were taken before neutron emission of the fissioned fragments.
}
\centering 
\begin{tabular}{|c|c|c|c|}
\hline 
 Method& Heavy Fragment  & Light Fragment   & Reference \\
\hline\hline
$^{240}$Pu(s.f) & 138.74$\pm$0.20 & 101.26$\pm$0.20&\cite{Thi81} \\
$^{240}$Pu(s.f) & 138.96$\pm$0.04 & 101.31$\pm$0.04&\cite{Wag84} \\
$^{239}$Pu($n_{th},f$)&139.67&100.33&\cite{Thi81} \\
$^{239}$Pu($n_{th},f$)&139.73$\pm 0.01$&100.27$\pm$0.01&\cite{Wag84}  \\
$^{240}$Pu($\gamma,f)$ (12 MeV)&139.88$\pm$0.14&100.12$\pm$0.14&\cite{Thi81} \\
$^{240}$Pu($\gamma,f)$ (15 MeV)&139.92$\pm$0.09&100.08$\pm$0.09&\cite{Thi81} \\
$^{240}$Pu($\gamma,f)$ (20 MeV)&139.84$\pm$0.08&100.16$\pm$0.08&\cite{Thi81} \\
$^{240}$Pu($\gamma,f)$ (30 MeV)&139.71$\pm$0.14&100.29$\pm$0.14&\cite{Thi81} \\
\hline 
\end{tabular}
\label{frag-mass-exp}
\end{table*}

These fragment masses can be compared directly to experimental data. Table \ref{frag-mass} displays the resulting fission fragment masses obtained from this theoretical study. We note that these can hardly be considered distributions, but rather should be taken as indications of what dynamical calculations can contribute to fission studies.
The two-fragment static configuration is included for comparison. It is important to stress that this differs from the deformation-induced fragments. It produces  more symmetric states, which brings it closer to experimental results of neutron-induced fission. The Table also includes the particle number rounded to the nearest integer. It would be of interest to project the individual fragments onto the particle number \cite{Sim10}, to obtain a mass number distribution, although this would not necessarily agree with experiments \cite{Scamps2015}.

Further, Table \ref{frag-mass-exp} contains experimental data taken from Refs. \cite{Wag84,Thi81}, listing the most likely masses of the fission fragments. The references study various fission processes in $^{240}$Pu, including spontaneous fission, thermal neutron-induced fission, and various energy photon-induced fission reactions. The spontaneous fission data has been included for completeness. 

\begin{figure}[t]
\begin{center} 
\includegraphics[width=0.6\linewidth]{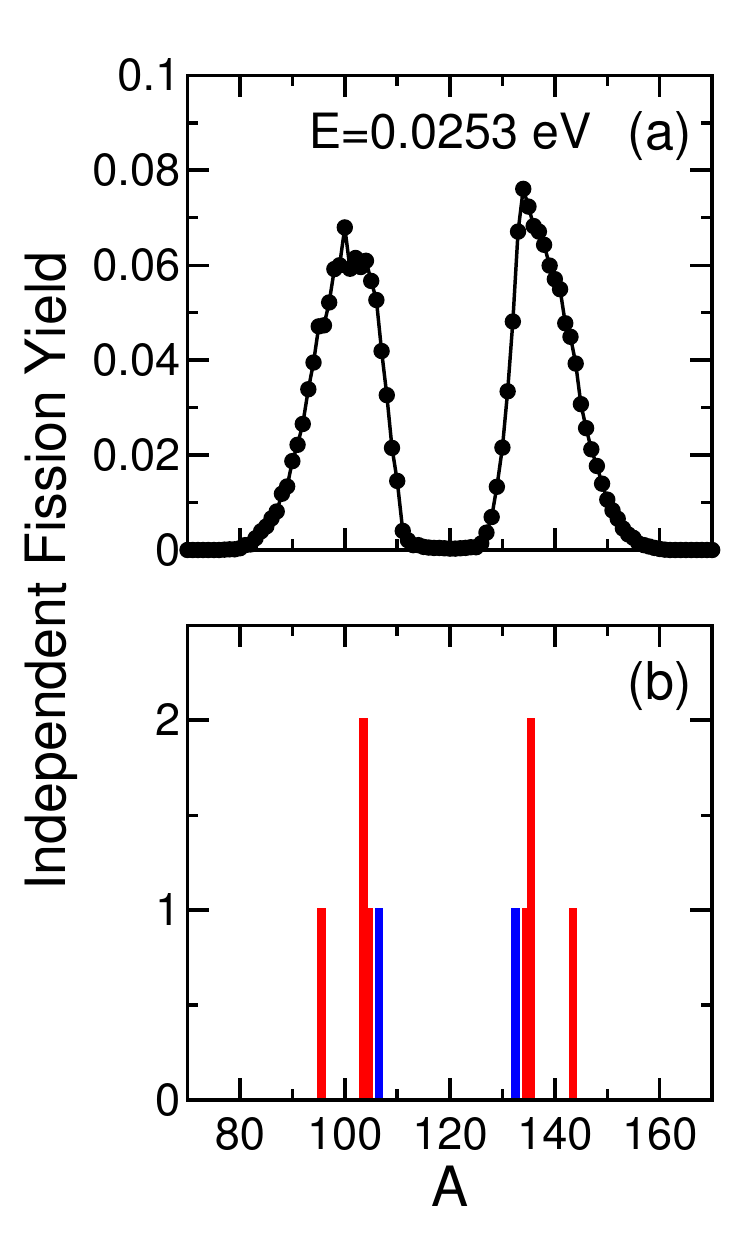} 
\caption{(Color online) (a) Experimental independent fission yields for neutron-induced fission  at $E=0.0253$ eV. The data are from Ref.~\cite{Jfis}. (b) Theoretical mass fragment distributions. The red bars correspond to the binned TDHF results and the blue bar corresponds to the static two-fragment mass split. See text for more details.}
\label{n-indiced}
\end{center}
\end{figure}

We emphasise that fission produces a range of masses. The values quoted in Table \ref{frag-mass-exp} correspond only to the most likely fissioned configuration. This is a crude comparison of experimental data to theoretical calculations. A more meaningful comparison would go through a characterization of the full mass distribution, in line with the recently proposed method of Ref.~\cite{Bertsch2015}. While we cannot at present produce a full mass distribution from projection into particle numbers, we can use each of the four different fission cases to build a schematic fission fragment mass distribution. Referring to Fig.\@ \ref{n-indiced}, which displays data for neutron-induced fission at low energies in panel (a), the obtained theoretical values in panel (b) fall well within the experimentally obtained mass distribution. With the limited data set available, the TDHF results seem to be consistent with the experimental data. Figure \ref{n-indiced} also includes for comparison the adiabatic, constrained static result of the two-fission pathway. All in all, dynamical effects seem to induce a larger mass asymmetry in the fission fragments.

\subsection{Energy of Fission Fragments}
\label{fragesec}

By applying masks around the spatial regions of the fission fragments, the energy density functional corresponding to each individual fragment may be obtained. We note, however, that the interpretation of the results is not as simple as in the two-fragment case. The nuclear part of the energy density functional is short-ranged. We therefore expect that an integral in the spatial region corresponding to the individual fragments will be a faithful representation of the nuclear part of the energy density. The Coulomb interaction, however, is long-ranged. As well as the Coulomb interaction within the individual fragments, there is a contribution from the interaction with one another. This contribution is missing in the present two-fragment integral. Further, the fragments decay by particle emission, which imparts some time dependence upon the integrated energy corresponding to the individual fragments. 

\begin{figure}[tbh]
\begin{center} 
\includegraphics[width=0.7\linewidth]{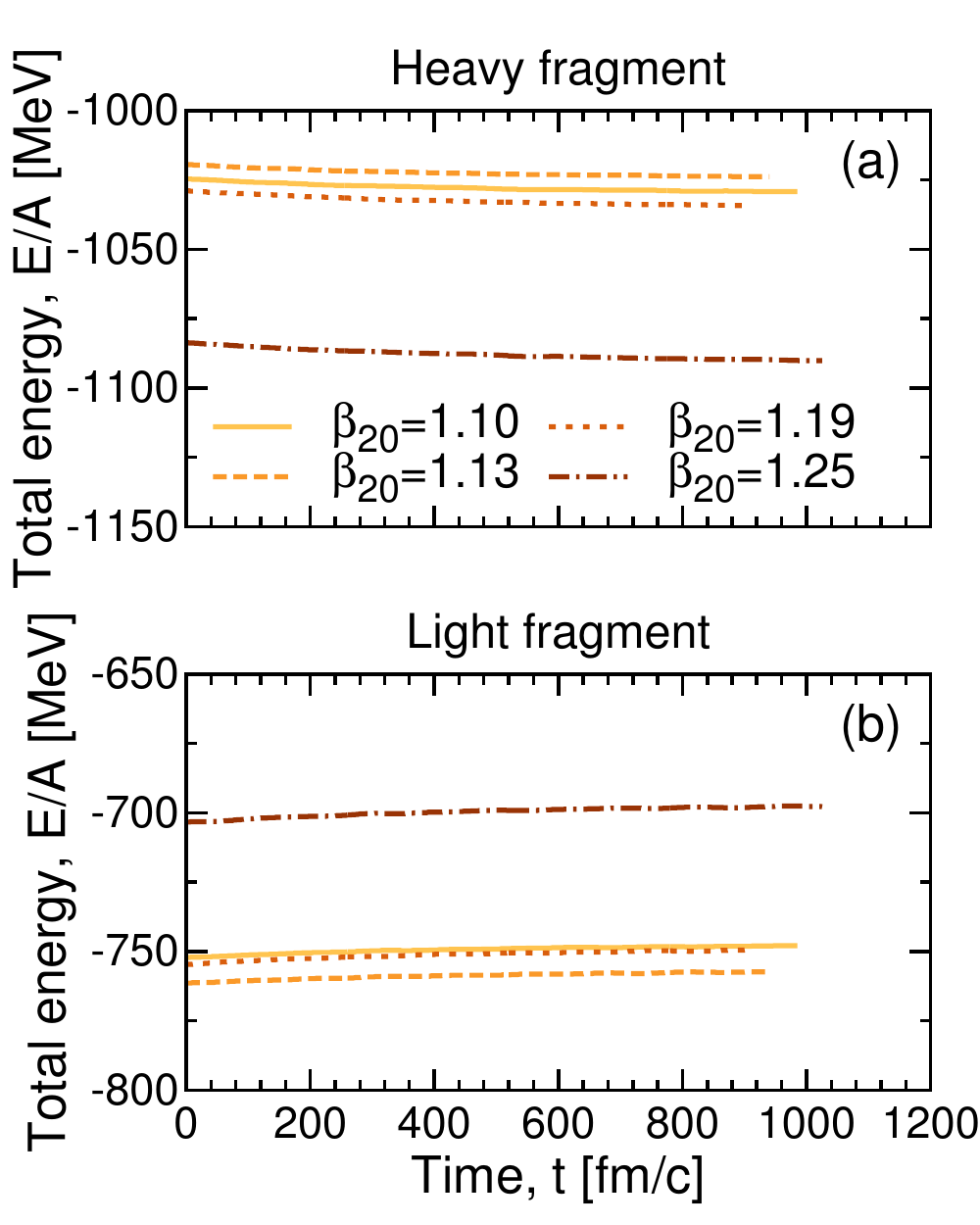} 
\caption{(Color online) 
Summed energy density functionals for the region of space corresponding to the (a) heavy and (b) light fission fragments. The quadrupole deformation of the initial state is labelled. The drift in the energy can be attributed mainly to the Coulomb interaction between the two fragments. See text for more details.}
\label{frag-edecomp}
\end{center}
\end{figure} 

The time evolution of the total integrated energy functional corresponding to the heavy and light (lower) fission fragments is shown in panels (a) and (b) of Fig.\@ \ref{frag-edecomp}, respectively. The time measurement at $t_{\text{sep}}=0$ begins when the fragments are sufficiently separated such that the masks no longer overlap. To a good approximation, the fragment total energies are constant over time. A slight drift is observed in the time evolution, owing to the long-range effects of the Coulomb interaction as well as particle decay. We only provide the total integrated energy of the post-fission fragments in Fig.\@ \ref{frag-edecomp}. The evolution of the decomposed terms corresponding to the individual fragments shows no remarkable behaviour. Hereafter the fragment energy at the cutoff time is denoted as $E^\ast$.

The total energies of the two fragments provide an alternative way of computing their collective kinetic energies. The total excitation energies of the fragments is the sum of their translational and internal collective kinetic energies. If we subtract the total integrated energy to the corresponding ground state energy, we obtain a new estimate for the fragment excitation energy.  This method complements the approach of interpolating the total collective kinetic energy of the system, as presented in Sec.\@ \ref{fissin}. As long as the ground state and the fragment energies are qualitatively correct, this second method should produce comparable results. 

The solver {\sc sky3d} has thus been applied to deduce the ground states of the fission fragments to the nearest integer particle number. The result for the ground-state energy of these isotopes is presented in column 4 of Table~\ref{gs-es-diff}. Here, it is debatable whether the energy functional in {\sc sky3d} contains all the terms required to calculate odd-odd and odd-even nuclei. The full time-odd contribution is presented in Ref. \cite{Suc11}, and the functional in {\sc sky3d} does not include all these terms. However, as the functional used for the static calculations is consistent with that applied to dynamic calculations, Galilean invariance is conserved. The functional used in {\sc sky3d} therefore satisfies all the invariance properties required to perform static calculations of odd-odd and odd-even nuclei, even if the functional is not in its most ``complete" form. We also note that pairing has been neglected in these calculations. The total pairing contribution to the energy functional is typically of the order of $0-10$ MeV, which is small compared to the excitation energies in the fissioning case. 

\begin{table*}[tbh]
\caption{Comparison of the fission fragment energies to the ground-state energy calculated using the SkM$^\ast$ interaction. The fragment total energy at the cutoff time is denoted by $E^\ast$ (see Fig.\@ \ref{frag-edecomp}), the ground state energy by $E_{gs}$, and the difference ($E^\ast-E_{gs}$) by $\Delta E$. See text for more details.} 
\centering 
\begin{tabular}{|c|c|c|c|c|c|c|c|c|} 
\hline 
 Static  $\beta_{20}$  &Heavy &$E^\ast$& $E_{gs}$  & $\Delta E$&Light  & $E^\ast$& $E_{gs}$&$\Delta E$  \\
 &Frag.&[MeV]&[MeV]&[MeV]&Frag.&[MeV]&[MeV]&[MeV] \\
\hline\hline 
1.10&$^{136}_{53}$I&-1029.22&-1118.31&89.09&$^{104}_{41}$Nb&-747.86&-854.54 &106.68 \\
1.13&$^{135}_{52}$Te&-1023.81&-1110.24&86.43&$^{105}_{42}$Mo&-757.38&-865.71&108.33 \\
1.19&$^{136}_{53}$I&-1034.23&-1118.31&84.04&$^{104}_{41}$Nb&-749.41&-854.54&105.13\\
1.25&$^{144}_{55}$Cs&-1090.17&-1162.47&72.30&$^{96}_{38}$Sr&-697.78& -796.94&99.16\\
\hline 
\end{tabular}
\label{gs-es-diff}
\end{table*}

\begin{figure}[tbh]
\begin{center} 
\includegraphics[width=\linewidth]{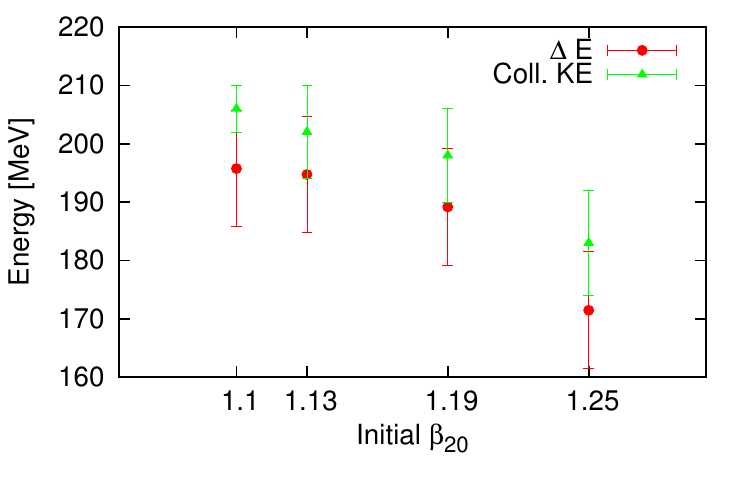} 
\caption{(Color online) Comparison of the summed $\Delta E$ from Table \ref{gs-es-diff} (solid circles) to the mean interpolated collective kinetic energy from Table \ref{frag-ke-tab}) (solid triangles) for each of the fissioning cases. Indicative error bars are displayed. See text for details. }
\label{frag-ke-ex-tab-comp}
\end{center}
\end{figure} 

We have access to two measurements of the total excitation energy of the systems, either by interpolating the evolution of the collective kinetic energy or by comparing the ground-state fragment energies to the excited fragment energies. Figure \ref{frag-ke-ex-tab-comp} displays the mean interpolated collective kinetic energies (circles) presented in Table \ref{frag-ke-tab}. We compare these to the total fragment excitation energy ($\Delta E_{\text{heavy frag.}} + \Delta E_{\text{light frag.}}$) for each fissioning case (triangles). The error bars in the values of $\Delta E$ display an uncertainty of $10$ MeV, a conservative estimate for typical pairing correlations, possible deformation energies, and also that only nearest-integer nuclei are considered. For the interpolated collective energy, the values presented are the mean of the three interpolations performed at different fragment separations. We present a preliminary, rough estimate of the error of this calculation, obtained from the standard deviation of these three data points (see Table \ref{frag-ke-tab}). 

Figure \ref{frag-ke-ex-tab-comp} shows that within the error bars, the results from the two techniques produce consistent values of the energy released in the fission process. A smaller error for both predictions could be achieved by performing the calculations up to the point where the Coulomb interaction is negligible, and obtaining ground-state energies incorporating pairing correlations. As mentioned, the energies presented here are measures of the \textit{total} excitation energy of the system. We note that this includes both excitation energy of the fragments and their relative motion, in contrast to the experimental definition that often neglects the latter. Section \ref{coll-ex-frag} will demonstrate a technique which may be used to decouple the translational kinetic energy from the internal collective excitation energy of the two separate fragments. These will provide further insight into the excitation mechanisms in TDHF fragment formation.

\begin{figure}[tbh]
\begin{center} 
\includegraphics[width=\linewidth]{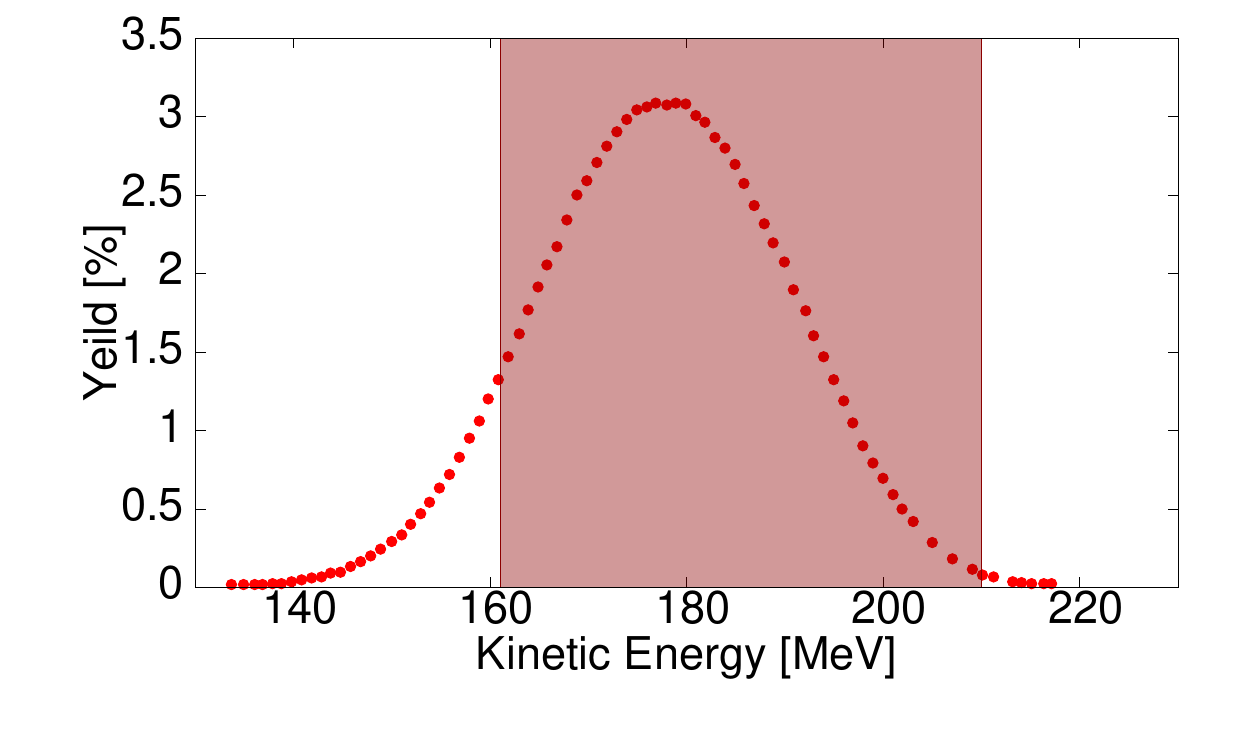} 
\caption{(Color online) Experimental kinetic energy distribution taken from Ref. \cite{Wag84}. The shaded box corresponds to the extremes of the range of kinetic energy values displayed in Fig.\@ \ref{frag-ke-ex-tab-comp}.}
\label{fiss-kin-exp}
\end{center}
\end{figure}

Figure \ref{fiss-kin-exp} shows experimental measurements of the kinetic energy reproduced from Ref. \cite{Wag84} for thermal neutron-induced fission in $^{240}$Pu. The range of collective kinetic energies deduced in this section are marked with a shaded box. By attributing the deduced total excitation energies solely to translational kinetic energy, this assumes that the internal collective excitation of the fragments are comparatively small. This will be demonstrated shortly. Despite the limited sample of theoretical data, the results agree  with the experimental range of values. In particular, our results fall well within the experimental distribution.

\subsection{Collective Excitation Modes of Fission Fragments}
\label{coll-ex-frag}

As mentioned, the excitation energy of the fission fragments is assumed to be dominated  by translational kinetic energy. However, as well as translational motion, the fragments undergo collective vibrations because of internal excitation. The collective excitation modes of the fragments may be investigated using both the time and the frequency domains. To perform Fourier analysis, we make use of the spectral power function 
\begin{equation}
P_\zeta(\omega) =  \left[ \text {Re } \zeta(\omega) \right]^2 + \left[ \text{Im } \zeta(\omega)  \right ]^2 \, ,
\end{equation}
where $\zeta(\omega)$ is the Fourier transform of the moment of interest. Unfortunately, owing to the limitations in the numerical grid size (the $z$ direction spanned 160 grid points, ranging from -79.5 to 79.5 fm), a signal corresponding to the evolution of the multipole parameters of the individual fragments could only be measured for approximately $1000$ fm/$c$ before the grid boundaries were approached. For a signal of this length, the resolution of the calculated power spectrum is of the order $\hbar\omega = \frac{\pi}{T_{\text{obs}}} \approx 0.6$ MeV \cite{Rei06a}. 

To extend the time evolution domain and consequently improve energy resolution, we adopted the following approach. Rather than performing the calculations in an impractically large numerical grid, a Galilean transformation is applied to each fission fragment. The boost momentum is chosen to cancel the corresponding linear momenta of each fragment. After boosting, the two fragments remain approximately still in the box, and their excitation modes can be studied for a much longer period. The evolution of the configuration with an initial $\beta_{20}=1.25$ is presented as an example of this new method. 

Inside the masked regions of space corresponding to the fragments, the linear momentum may be calculated by integrating the current density:
\begin{equation}
\bm p_\text{frag} = \int d \bm r \, \, \bm j(\bm r)  \,\,\, .
\end{equation}
This momentum therefore has units of velocity \cite{Mar13}. The linear momentum of the fragments may then be instantaneously removed by applying a Galilean boost to the single-particle wave functions,
\begin{equation}
\bar \varphi(\bm r) = \exp 
\left( i \frac{\bm p_\text{frag} \cdot \bm r }{ A_\text{frag} } \right) \,\varphi (\bm r) \, ,
\end{equation}
where $A_\text{frag}$ is the integrated particle density corresponding to the fragment and $\varphi(\bm r)$ are the single-particle wave functions. The Galilean transformation should be applied in the masked region of space with the corresponding momentum for each fragment. The effect of the transformation is to effectively boost the particles in the opposite direction with the exact momentum they are propagating with through the grid.

\begin{figure}[tbh]
\begin{center} 
\includegraphics[width=\linewidth]{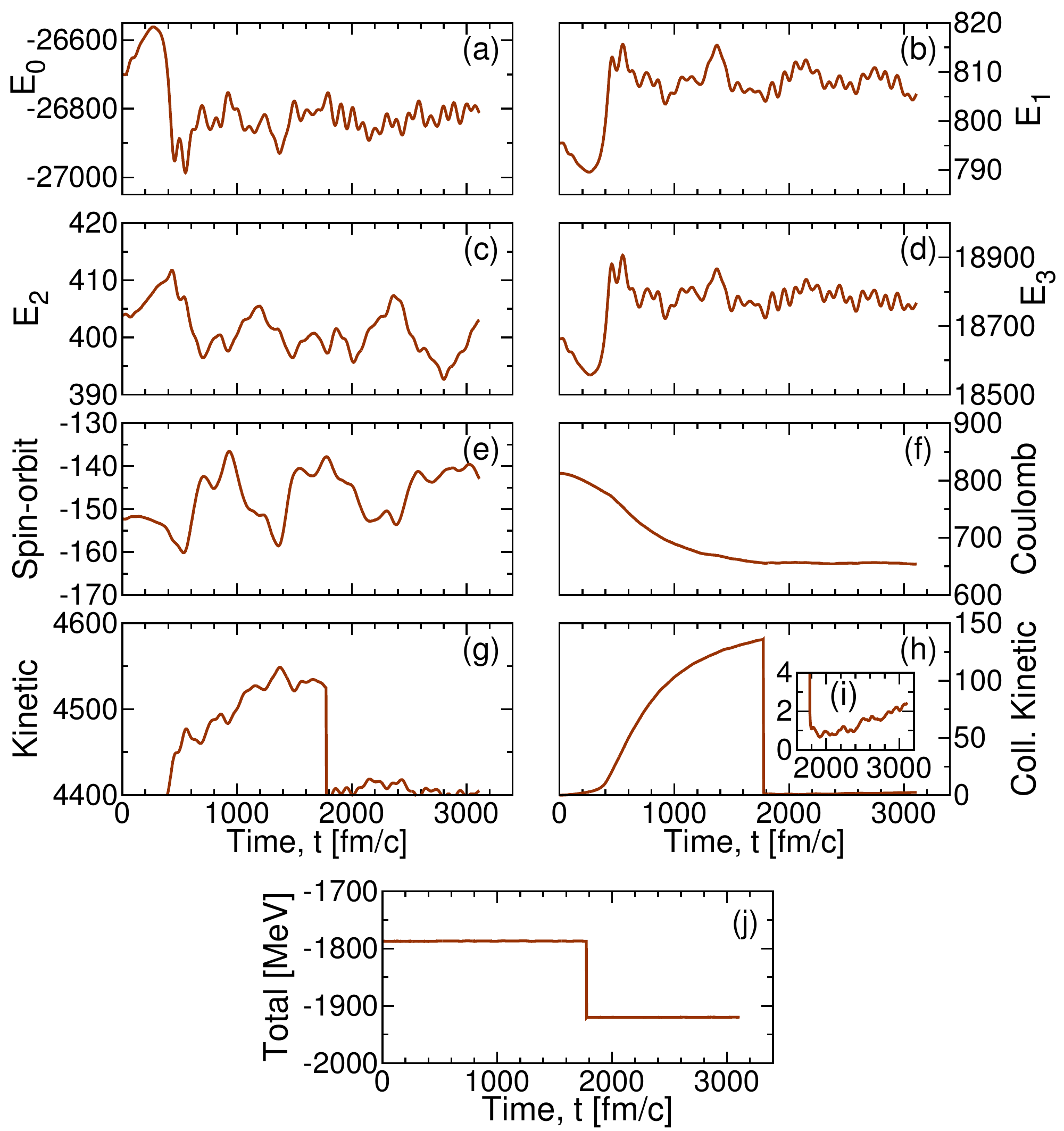} 
\caption{(Color online) Decomposed contributions to the energy of the system for the case with initial deformation $\beta_{20}=1.25$. A Galilean transformation has been applied to remove the linear momentum of the individual fragments once the separation between the fragment center of mass reaches $100$ fm. The calculation is terminated once the separation exceeds $105$ fm. }
\label{gal-trans}
\end{center}
\end{figure} 

Figure \ref{gal-trans} shows the decomposed energy functional for the fissioning case with initial $\beta_{20}=1.25$. The Galilean transform was applied when the separation of the fragments reached $100$ fm (corresponding to $t \approx 1900$ fm/$c$), and the calculation terminated at separation $105$ fm. Upon application of the transformation, panel (j), corresponding to the total energy, displays a decrease of approximately $140$ MeV. This mirrors the drop in collective kinetic energy by the same amount, which corresponds to an instantaneous removal of excitation energy owing to translational motion. This demonstrates that the total excitation energy is dominated by contributions from translational motion, rather than internal collective excitations. As for the other energy contributions, the nuclear and Coulomb energies remain unaltered before and after the boost is applied around $t \approx 1900$ fm/$c$. This is to be expected, because these terms are all Galilean invariant.

The collective kinetic energy drops instantaneously to $\approx1.1$ MeV following the transformation, as seen in inset panel (i). Because the translational energy has been removed at this point, the remaining collective energy can be interpreted as the sum of the internal excitation energy shared between the two fragments. Reference \cite{Sim13} discusses an alternative method to deduce the internal collective excitation energy of the fragments, but the method applied assumes \textit{a priori} knowledge of the fission products of the system. 

The internal collective excitation energy is very small compared to the total excitation energy released in the fission process, which is $\approx180$ MeV (see Fig.\@ \ref{frag-ke-ex-tab-comp}, $\beta_{20}=1.25$). This justifies the previous assumption that the final collective excitation energy deduced in TDHF is dominantly translational kinetic energy, so it may therefore be compared to the experimentally measured kinetic energies (Fig.\@ \ref{fiss-kin-exp}). Further, the energy functional contributions (Fig.\@ \ref{gal-trans}) may be compared to these in Fig.\@ \ref{fiss-edf}, where the calculation was terminated at the point where the transformation is applied in this case. Figure \ref{gal-trans} demonstrates that the nuclear potential part of the energy functional is unaffected by the transformation. The calculation was performed in a grid of identical dimensions to those presented in Fig.\@~\ref{fiss-edf} ($48\times48\times160$ points), and the time elapsed has effectively doubled from those previous calculations. As the measurement time of the post-fission fragments has been elongated, the resolution of the resulting power spectra will be enhanced accordingly.

We note that the Coulomb interaction is long-ranged, so that even at a separation of $100$ fm  there is an interaction between the fragments. Translational motion resumes after the Galilean transformation is applied, and the translational kinetic energy slowly increases. This can be seen by the gradual increase of the collective kinetic energy in Fig.\@ \ref{gal-trans} following the transformation. Therefore, the center-of-mass separation eventually reaches $105$ fm and the calculation is terminated. We note that, in principle, one could reapply the Galilean transformation at every iteration to extend the total time and improve energy resolution. This artificially prolonged Coulomb interaction between the fragments does not alter significantly the dynamics of the fragments. Any potential adverse effects have not been studied in detail, and alternate methods can also be applied to mitigate this increased Coulomb interaction \cite{Flocard1981,Simenel2004}.

\begin{figure}[h!]
\begin{center} 
\includegraphics[width=0.6\linewidth]{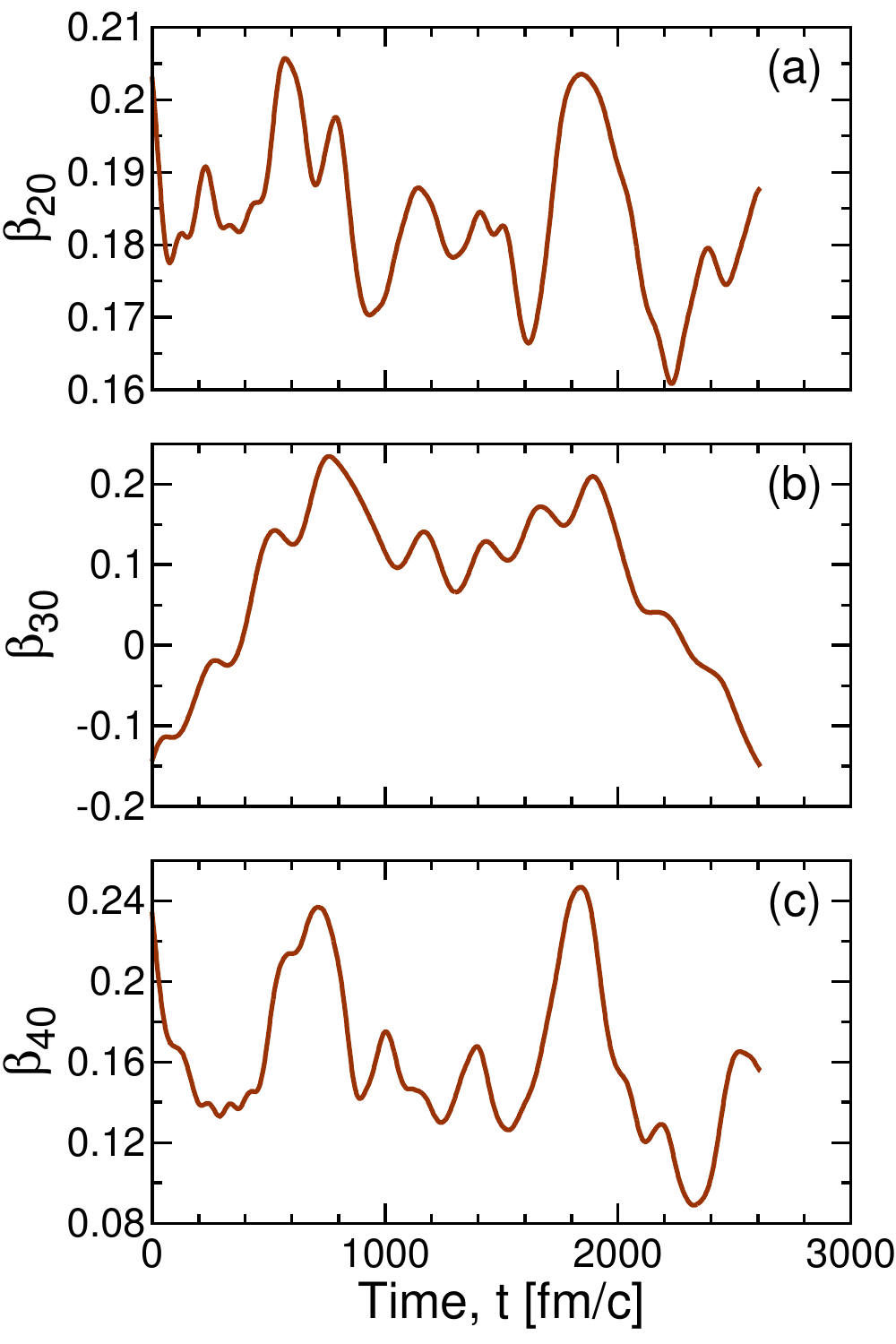} 
\caption{(Color online) Evolution of (a) quadrupole, (b) octupole and (c) hexadecapole deformation parameters for the heavy fission fragment. The initial deformation was $\beta_{20}=1.25$. The measurement time is significantly extended by applying the Galilean transformation to remove the linear momentum of the fragments. See text for more details. }
\label{gal-upper}
\end{center}
\end{figure} 

\begin{figure}[h!]
\begin{center} 
\includegraphics[width=0.6\linewidth]{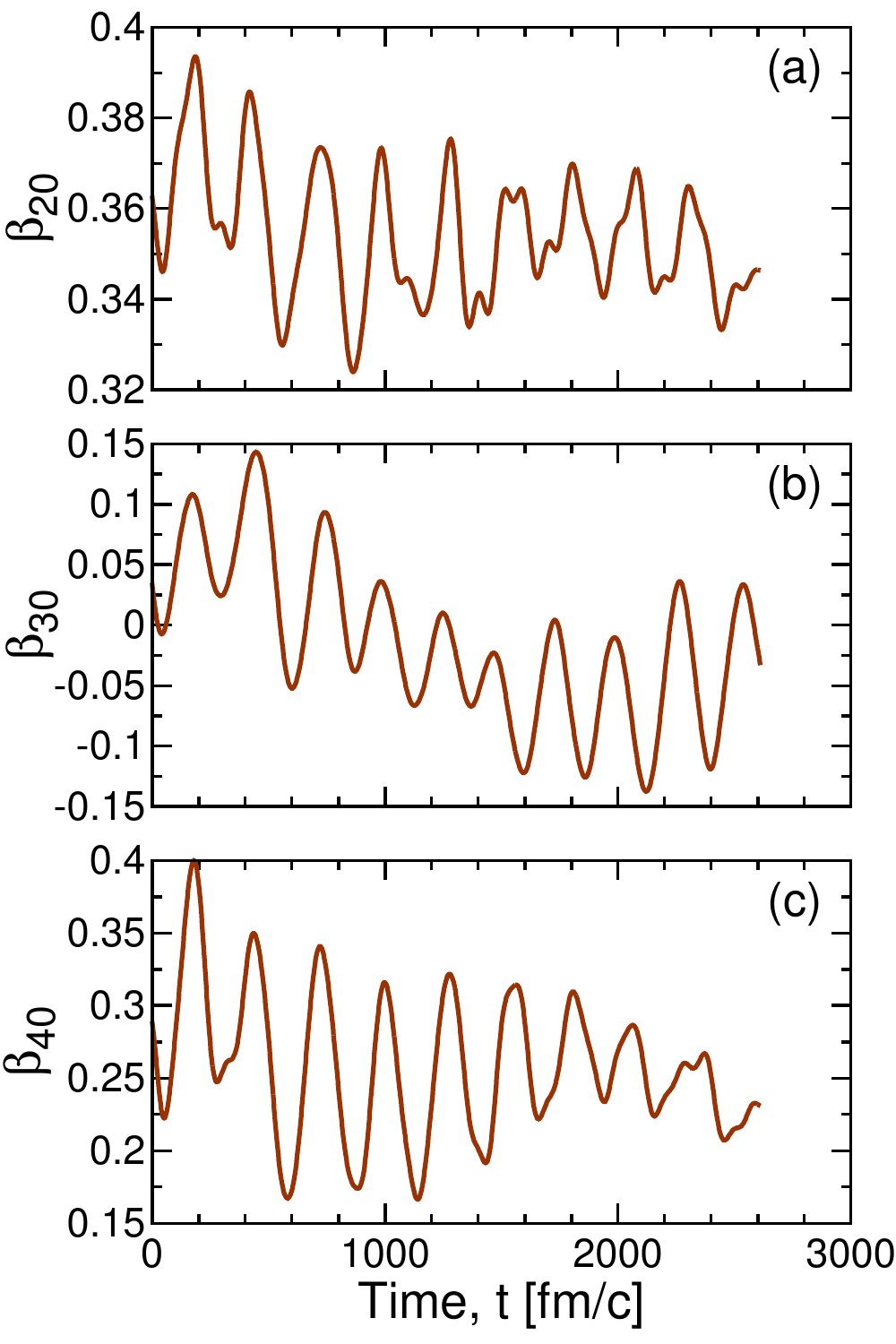} 
\caption{(Color online) Evolution of (a) quadrupole, (b) octupole and (c) hexadecapole deformation  parameters for the light fission fragment. The initial deformation is $\beta_{20}=1.25$. }
\label{gal-lower}
\end{center}
\end{figure} 

\begin{figure}[h!]
\begin{center} 
\includegraphics[width=0.6\linewidth]{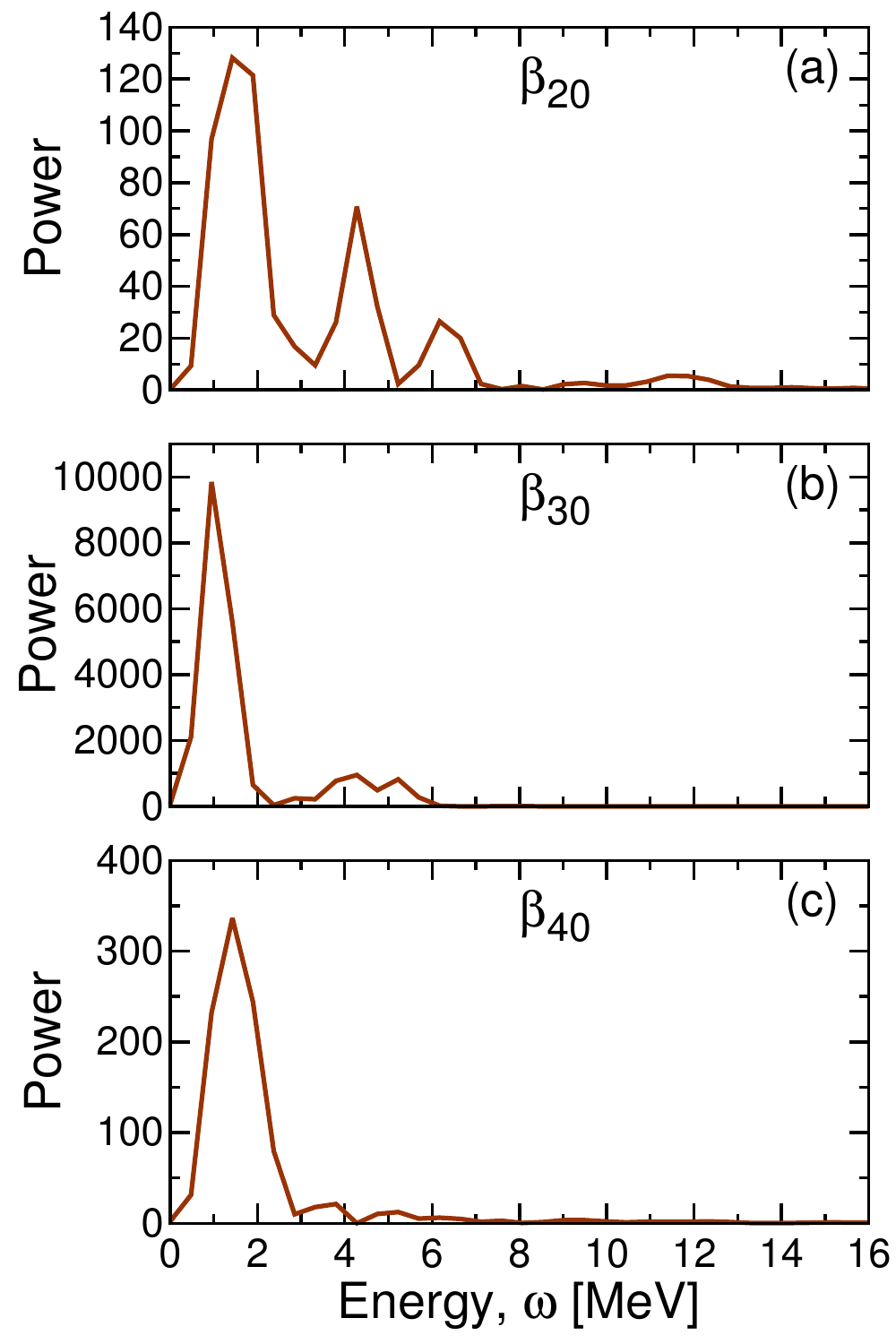} 
\caption{(Color online) Power spectra corresponding to Fig.\@ \ref{gal-upper} (heavy fission fragment). The resolution is significantly improved due to the longer measurement time available with the use of Galilean transformations to remove the linear momentum of the fragments.}
\label{pow-upper}
\end{center}
\end{figure}

\begin{figure}[h!]
\begin{center} 
\includegraphics[width=0.6\linewidth]{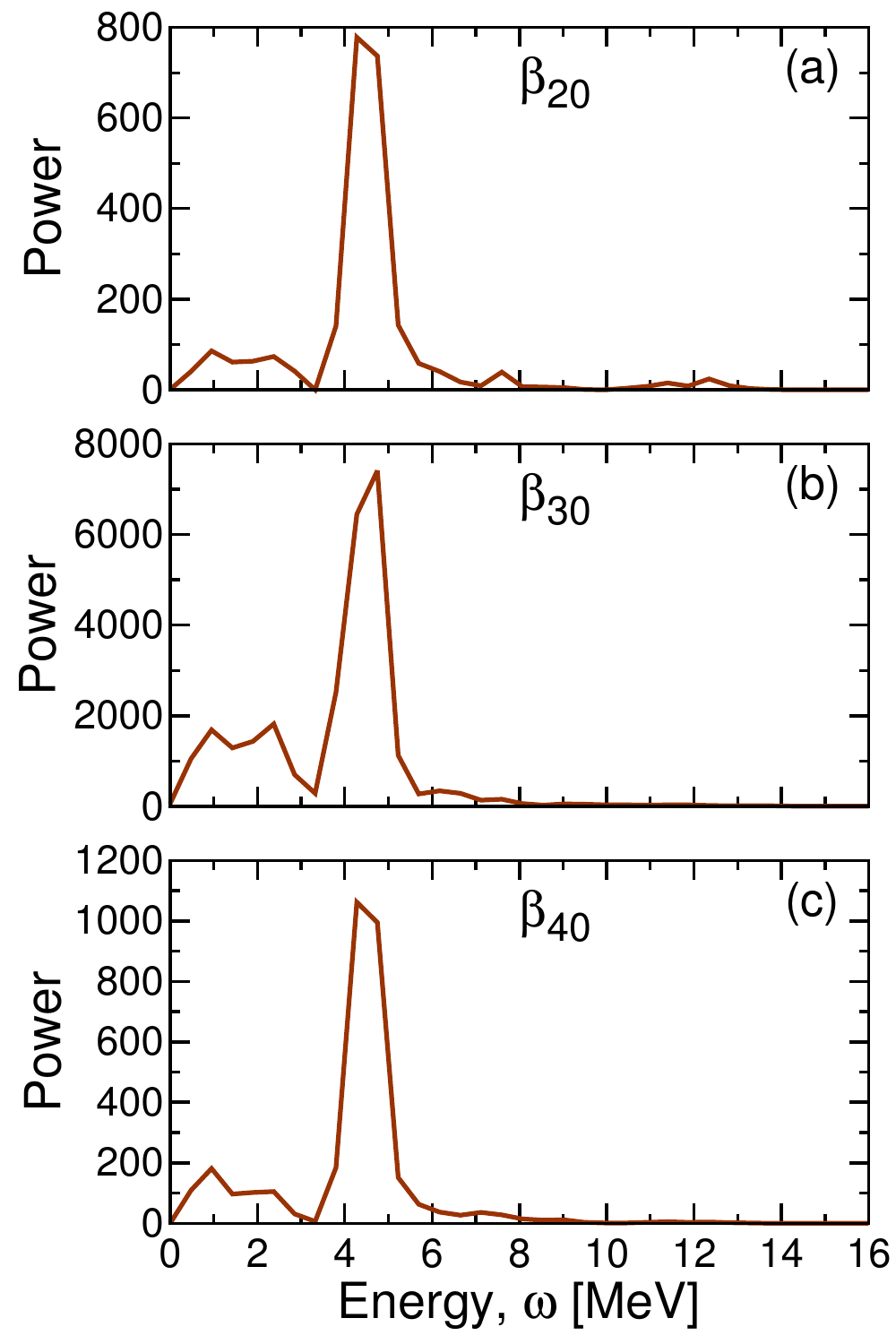} 
\caption{(Color online) Power spectra corresponding to Fig.\@ \ref{gal-lower} (light fission fragment).}
\label{pow-lower}
\end{center}
\end{figure}

The evolution of the multipole deformation parameters for the two fragments are shown in Figs. \ref{gal-upper} and \ref{gal-lower}, corresponding to the heavy and light fragment, respectively. The corresponding power spectra are shown in Figs. \ref{pow-upper} and \ref{pow-lower}. Let us stress that the resolution of the spectra is $\approx 0.4$ MeV, a significant improvement with respect to what would be obtained without extending the measurement time.

The evolution of the multipole fragments is qualitatively similar for both fragments. The quadrupole oscillations are centered around values of $\beta_{20} \approx 0.18$ and $0.35$ for the heavy and light fragments, respectively. The corresponding hexadecapole moments oscillate in phase with the quadrupole deformation around central, non-zero values. The corresponding octupole deformations, in contrast, are modulated around a zero value, and it is difficult to ascribe a final octupole deformation for these (excited) fragments. While the specific deformations might not be particularly relevant, the excitation patterns of the multipoles provide information on the collective vibrations of both fragments. We note that, whereas a pattern of well defined, relatively rapid oscillations are found for the light fragment, the heavy fragment multipoles have a more erratic time evolution.  

These features are reflected in the corresponding power spectra. We insist here that these can be obtained only with the required resolution after a boost has been performed on both fragments. Within the resulting spectra presented for the heavy fragment in Fig.\@ \ref{pow-upper}, there is a well-defined peak for each multipole parameter between $\approx 1$ and $3$ MeV. For the light fragment, in contrast, Fig.\@ \ref{pow-lower} shows a very well-defined peak for each multipole parameter between $4$ and $6$ MeV. The latter reflects the very well-defined oscillations in the moments observed earlier. We speculate that these well-defined oscillations correspond to collective excitations within the two fragments. It would be of interest to compare the excitation modes obtained for the post-fission fragments to standard calculations of resonances in the corresponding nuclei. This would give access to a microscopic understanding of the fragment excitation properties, including their temperatures and phonon structure.

\section{Conclusions}
\label{sec:conclusion}

We have presented an analysis of the fission process using TDHF techniques as implemented in {\sc sky3d}. Starting from the calculated one-fragment quadrupole-constrained PES, the dynamics of fission were investigated. Deformation-induced fission was explored by releasing the quadrupole constraint and time-evolving a selection of states situated below, around and beyond the second static fission barrier. Three behaviours were observed. The states with a quadrupole deformation below the peak of the fission barrier undergo vibrations corresponding to a collective giant resonance. For these states, DIF is forbidden in TDHF, because a collective tunnelling through the barrier must occur to reach a fissioned configuration.

A different behaviour is observed for the evolution of states which are situated beyond the peak of the second static fission barrier, but before the critical point where the static one and two-fragment pathways intercept. Upon time evolution up to $9000$ fm/$c$, these states also fail to fission, but the dynamics are not typical of collective giant resonant modes. The repulsive Coulomb force attempts to drive the configuration towards a fission point, but owing to the competition with the attractive terms in the energy functional, scission does not occur. DIF is inhibited for these initial configurations, and it can only be speculated if these states would eventually fission with a longer time evolution.

For states with a static deformation exceeding the intersection of the one and two-fragment fission pathways, DIF was observed upon time evolution. We interpret that, because a static two-fragment configuration exists with greater binding energy than the one-fragment configuration, it becomes energetically possible for the one-fragment configurations to evolve to fission with only a modest rearrangement of the densities. The evolution of the pre-fission fragment displays a rearrangement of the densities up until around the point of scission. At this point, the Coulomb repulsion between the pre-formed fragments overpowers the nuclear potential, and translational motion sets in as the fission products rapidly accelerate away from one another. The timescale for DIF varies depending upon the deformation of the initial state. The least deformed configuration demonstrates a density rearrangement phase lasting approximately $1500$ fm/$c$, whereas the most deformed configuration is initially close to the point of scission and the neck ruptures within $100-200$ fm/$c$.

A selection of fission products was observed for the various initial configurations considered. When compared to experimental measurements of neutron-induced fission processes, the agreement of the calculated fragment masses demonstrates promising results, although several effects, particularly dynamical pairing, are missing in our approach. The energy released is shown to be dominantly translational kinetic energy, and agreement between theory and experiment was found to be reasonable when comparing the calculated and measured kinetic energies of the post-fissioned systems. We have pioneered a method to remove translational kinetic energy of the post-scission fragments to provide details of their internal excitations. 

This initial investigation into fission induced by deformation effects using time-dependent techniques has provided insightful results on the interplay between structure and dynamics in the fission process. We have also devised a set of useful computational analysis tools for the post-scission fragments. Deformation-induced fission, however, provides a limited amount of two-fragment configurations. Further extensions of the present research involving particle-number projection could provide access to more relevant mass distributions. Moreover, a refined step-by-step linear momentum removal could help extend the time length and energy resolution of the two-fragment excitation data. 

The present analysis is relevant for a variety of fission processes. Spontaneous fission would presumably tunnel the system across the PES barrier into random states within the ``allowed" region. These would subsequently decay into the two-fission pathway. One can in principle explore a wider landscape of fission fragments by exciting nuclei along their fission paths \cite{Goddard14}. This theoretical approach can be linked more naturally to induced fission, where the energy deposited by external probes induces the fission process. We plan to explore this approach using TDHF techniques in the near future. 

\begin{acknowledgments}
This work is supported by STFC, through Grants ST/I005528/1, ST/J000051/1 and ST/J500768/1. This research made use of the STFC DiRAC HPC cluster.  
\end{acknowledgments}

\bibliographystyle{apsrev4-1}
\bibliography{biblio}

\end{document}